\documentclass[preprint]{elsarticle}

\makeatletter
\def\ps@pprintTitle{%
 \let\@oddhead\@empty
 \let\@evenhead\@empty
 \def\@oddfoot{}%
 \let\@evenfoot\@oddfoot}
\makeatother

\usepackage{amsmath}
\usepackage{url}
\usepackage{bm} 
\usepackage{graphicx}
\usepackage{color}
\usepackage{amssymb} 
\usepackage{amsfonts}
\usepackage{multirow}
\usepackage[export]{adjustbox}

\newcommand{\field}[1]{\mathbb{#1}}   

\DeclareMathAlphabet{\mathscr}{OT1}{pzc}%
                                 {m}{it}

\begin{document}

\title{Statistical mechanics of ontology based annotations}

\author{David C. Hoyle}

\address{Thorpe Informatics Ltd., \\
Adamson House,\\
Towers Business Park, \\
Wilmslow Rd., Manchester, \\
M20 2YY, UK.}

\ead{david.hoyle@thorpeinformatics.co.uk}

\author{Andrew Brass}

\address{School of Computer Science, \\
University of Manchester,\\
Kilburn Building, \\
Oxford Rd., Manchester, \\
M13 9PL, UK.}

\ead{a.brass@manchester.ac.uk}

\begin{abstract}
We present a statistical mechanical theory of the process of annotating an object with terms selected from an ontology. The term selection process is formulated as an ideal lattice gas model, but in a highly structured inhomogeneous field. The model enables us to explain patterns recently observed in real-world annotation data sets, in terms of the underlying graph structure of the ontology. By relating the external field strengths to the information content of each node in the ontology graph, the statistical mechanical model also allows us to propose a number of practical metrics for assessing the quality of both the ontology, and the annotations that arise from its use. Using the statistical mechanical formalism we also study an ensemble of ontologies of differing size and complexity; an analysis not readily performed using real data alone. Focusing on regular tree ontology graphs we uncover a rich set of scaling laws describing the growth in the optimal ontology size as the number of objects being annotated increases. In doing so we provide a further possible measure for assessment of ontologies.
\end{abstract}
\maketitle

\vspace{2pc}

\let\thefootnote\relax\footnote{Published in {\it Physica A}, \url{http://dx.doi.org/10.1016/j.physa.2015.09.020}}
\let\thefootnote\relax\footnote{\copyright 2016. This manuscript version is made available under the CC-BY-NC-ND 4.0 license \url{http://creativecommons.org/licenses/by-nc-nd/4.0/}}

\section{Introduction \label{sec:Intro}}
With larger and more complex data sets becoming increasingly common, the annotation of data in order to semantically enrich it is a crucial task within data science \cite{TaylorJoudrey2008, Davies2009}. For example, in molecular biology a gene can be annotated by domain experts with terms, $t$, from a controlled vocabulary, thereby allowing other researchers to comprehend the function and role of that gene. Similarly, user tagging of information sources such as documents, photographs, or online content, provide additional meta-data and lead to emergent but uncontrolled vocabularies (often called folksonomies \cite{PetersStock2007}). Terms within a vocabulary can be further organized in a hierarchical structure such as a taxonomy \cite{TaylorJoudrey2008}, in which terms closer to the root of the hierarchy are less specific than those further from the root. Hierarchical organization of vocabulary terms can also be used to specify richer semantic relationships between terms - richer than just indicating that one term is a simply a sub-type of another. These richer hierarchical semantic structures are generally called ontologies \cite{Gruber2008}.

\begin{figure}[t]
\centering
\includegraphics[width=5.45cm, keepaspectratio=true, valign=t, trim=0cm 0cm 0cm 2.00cm bb=0 0 792 612]{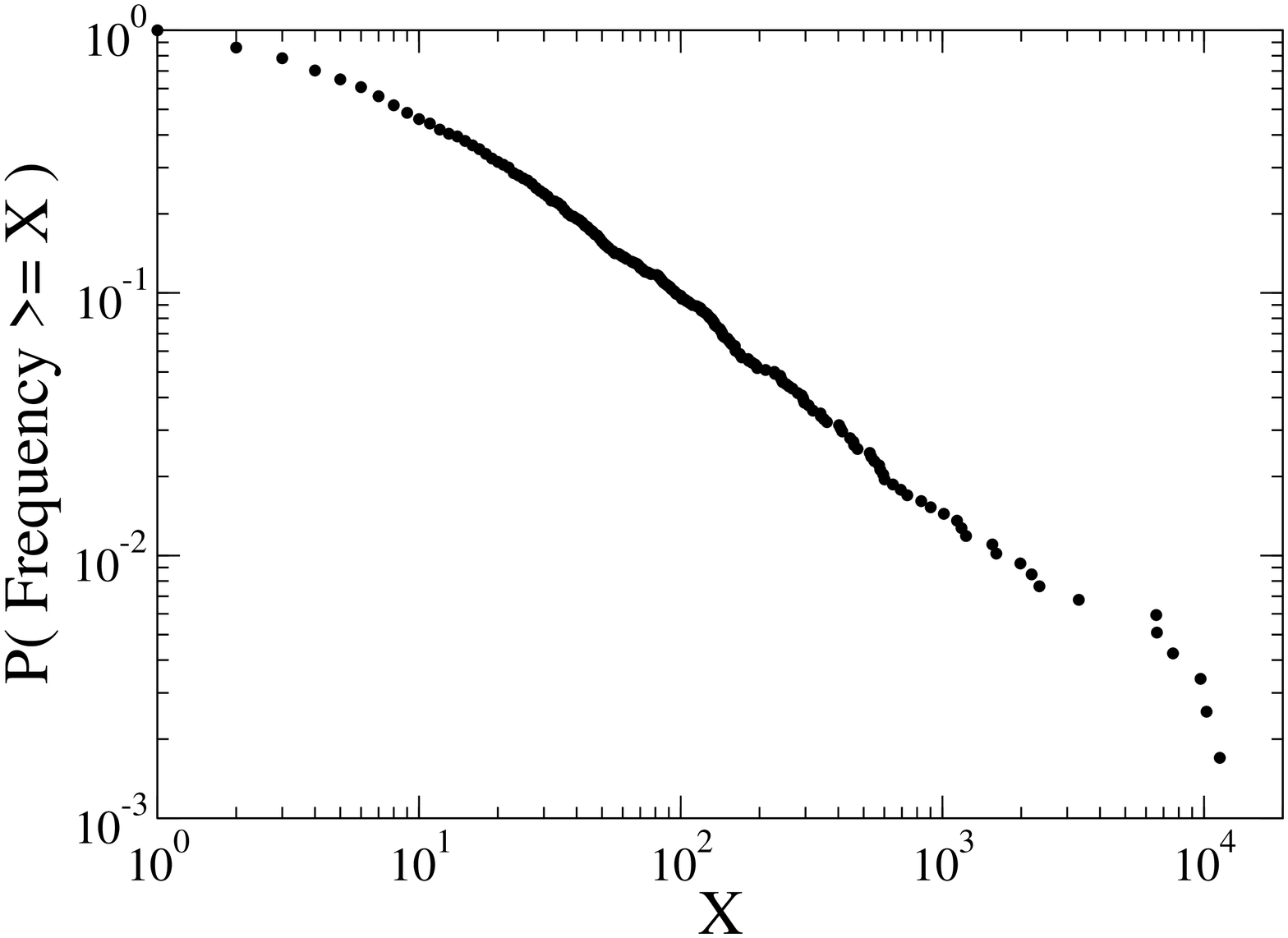}
\includegraphics[height=3.25cm, valign=t, bb=0 0 641 321]{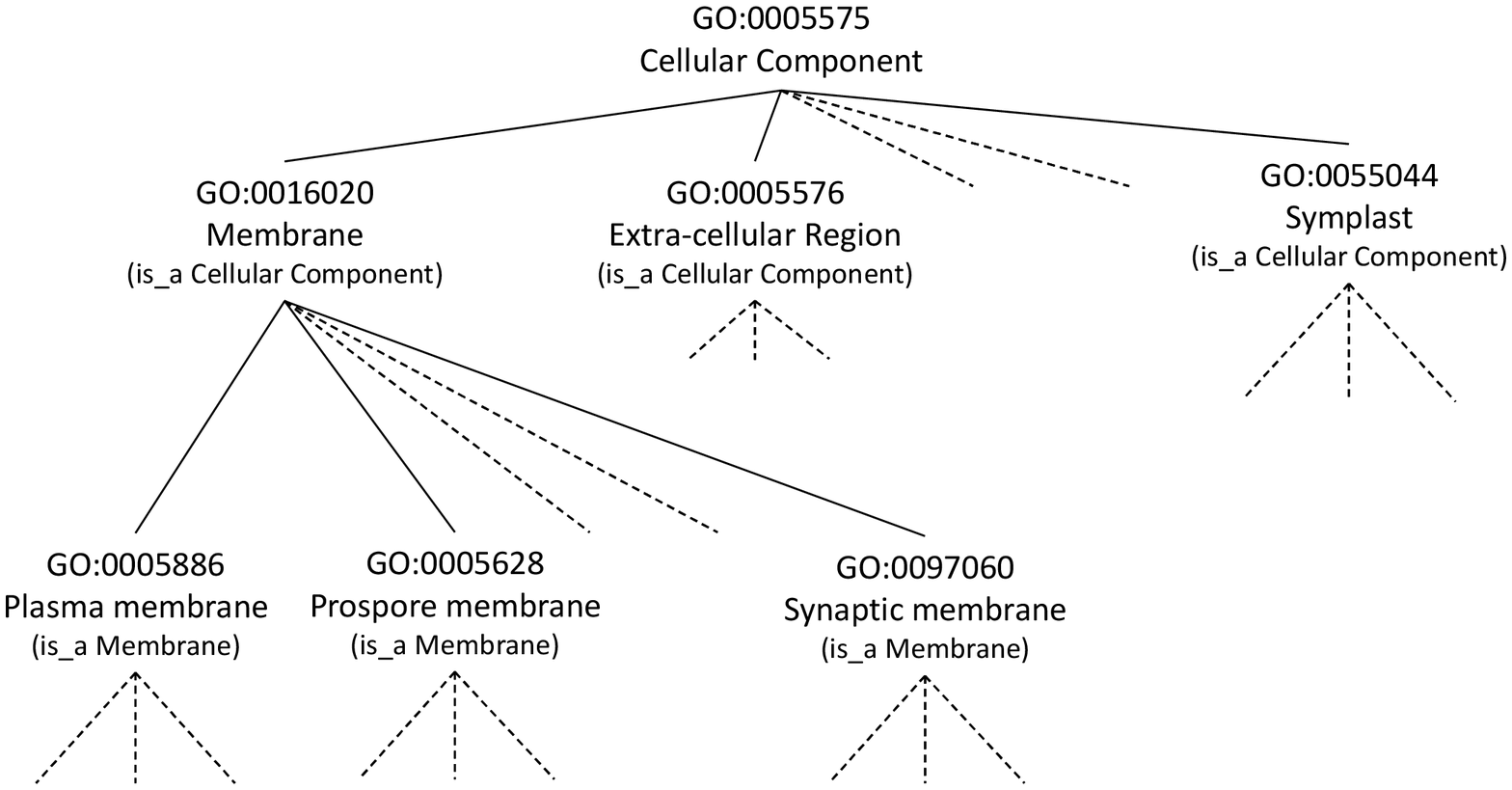}
\caption{Left-hand plot shows a Zipf's law plot for GO gene annotations contained within
Human GOA. Annotations have been taken from the cellular component sub-ontology. Terms near the root of the cellular component of GO are shown, for illustration, in the right-hand schematic - the dashed lines indicate the presence of further child terms.}
\label{fig:KalankeshExample}
\end{figure}

The annotations that result from an ontology or taxonomy can exhibit interesting patterns. For example, Kalankesh {\it et al.} \cite{Kalankesh2012} have shown that distributions of term frequencies, $f_{t}$, taken from Gene Ontology (GO) \cite{Ashburner2000} annotations typically follow Zipf's law \cite{Zipf1936, Zipf1949}. Figure \ref{fig:KalankeshExample} shows a Zipf's law plot for annotations taken from the cellular component sub-ontology of GO. The schematic on the right-hand side of Fig.\ref{fig:KalankeshExample} shows, for illustration, part of the ontology that was used to produce the annotation data set plotted on the left-hand side of Fig.\ref{fig:KalankeshExample}. Statistical mechanics provides us with a natural tool to understand these annotation patterns, by allowing us to develop a formalism that quantifies both the natural variations in the annotation process, and the ontology structure itself. Although structure-based measures of ontologies already exist \cite{Yao2011}, within this work we are quantifying the ontology structure from the perspective of the annotations that arise, rather than simply quantifying the ontology structure in isolation. The goals, and ultimately the benefits, of developing a statistical mechanics based formalism are both practical and theoretical.

\subsection{Ontologies as information stores}
Ontologies and taxonomies, whether formally constructed or emergent, represent a store of information. Organizing a hierarchical store of information requires effort to be expended to create an ordered structure.  Work by Ferrer i Cancho {\it et al.}, within an information theoretic framework, has shown how heavy tailed and essentially hierarchical patterns of term usage can arise simply from a principle of minimizing the communication effort expended when using those terms \cite{iCancho2003, iCancho2005, iCancho2007}. Within this current paper we also use information theoretic ideas, but it is the process of transferring information from an ontology to an annotated object that we study, {\it i.e.}, {\it after} the hierarchical term structure has been determined or prescribed. We do so using an explicit statistical mechanical model that takes into the structure of the ontology. Whilst existing work within the literature has used a specific Hamiltonian to study patterns of word usage, that work has not {\it per se} been interested in the impact of any underlying prescribed structure in the vocabulary \cite{Kosmidis2006}. Similarly, novel work by Palla {\it et al.} \cite{Palla2008} and Tib\'{e}ly {\it et al.} \cite{Tibely2012} has related tag usage patterns to ontology structure, but focused on an in depth study of observed tag patterns, rather than taking a Hamiltonian model based approach.

The remainder of this paper is organized as follows - in Section \ref{sec:Theory} we express the annotation process as an ideal lattice gas model in an inhomogeneous field. In Section \ref{sec:LeafDesc} we use the lattice gas model to understand the term frequency patterns seen by Kalankesh {\it et al.} \cite{Kalankesh2012}, and we identify, $LD_{t}$, the number of leaf descendants of a node $t$, as the key quantity controlling the expected term usage frequencies. In Section \ref{sec:MaxEntField} we derive the most likely natural form for the inhomogeneous field strength, thereby giving rise to a local measure of the ontology. This natural form for the inhomogeneous field also allows us, in Section \ref{sec:Metric}, to construct an ensemble of ontologies of differing complexity. By restricting the ensemble to the class of regular trees we reveal in Section \ref{sec:regTrees} a set of transitions in the optimal tree size, and associated scaling laws, as the number of objects being annotated is increased. Finally in Section \ref{sec:Conc} we discuss a number of possible extensions of the statistical mechanical approach to quantifying ontology structures.

\section{Statistical mechanical theory of the annotation process \label{sec:Theory}}
\begin{figure}[htb]
\begin{center}
\scalebox{0.83}{%
\includegraphics*[bb = 125 600 335 720]{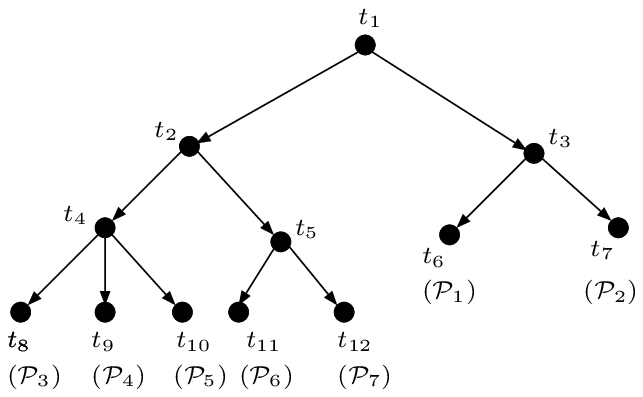}
\includegraphics*[bb = 135 595 345 720]{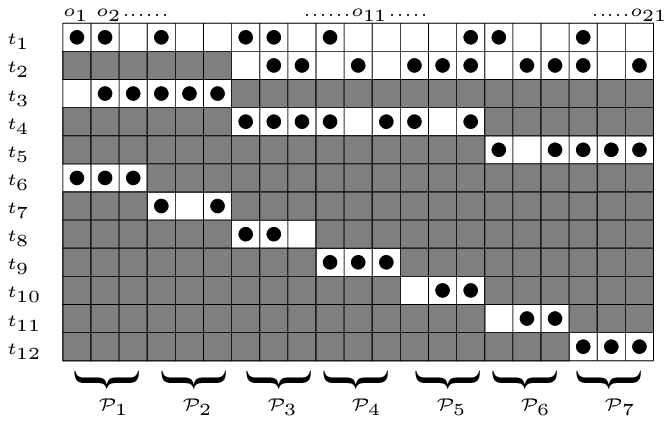}}
\end{center}
\caption{Left hand schematic shows an example (tree) DAG representing an ontology. The nodes represent terms, denoted by $t_{1}-t_{12}$, which may be selected to annotate objects that belong to classes represented by the leaves of the DAG. Each leaf corresponds to a unique set of paths from the root, and so we denote the classes by ${\mathcal P}_{1}-{\mathcal P}_{7}$. The directed edges of the DAG indicate semantic relationships between the terms. The right-hand side shows a lattice-gas schematic of an example annotation, {\it i.e.} a selection of terms appropriate to 21 objects, $o_{1}, o_{2},\ldots, o_{21}$, that have been assigned to the 7 classes ${\mathcal P}_{1}-{\mathcal P}_{7}$ (3 objects per class). A cell occupied by a particle in the lattice-gas schematic indicates the term has been selected to annotate that object, whilst a filled (grey) cell indicates a cell that cannot be occupied as a consequence of the ontology structure.}
\label{fig:OntologySchematics}
\end{figure}

We consider an ontology to be represented by a rooted Directed Acyclic Graph (DAG) \cite{BangJensen2009, Cormen2009}. An example DAG, in this case a tree, is shown in Figure \ref{fig:OntologySchematics}. Real-world ontologies are typically not pure trees, and we use a tree structure simply for illustrative purposes. The formalism we develop in this section will be equally applicable to any valid DAG structure. Associated with each node of the DAG is a particular term, and we use node and term interchangeably.

The directed edges between nodes of the DAG indicates that the child node conveys a possibly more specific meaning than the parent node, or represents a more specific subset of objects. For such an increase in specificity to occur the ontology topology must be ultimately be tree-like. Clearly, if a particular term is applicable to an object then so are the terms associated with any of the ancestor nodes.

At a leaf, we have a term that is applicable to a small subset of highly defined objects, possibly even a single object. Each leaf of the DAG determines a unique path or set of paths, denoted by ${\mathcal P}$, consisting of all terms that can be traversed in moving from the root to the leaf. Thus we use leaf, class and ${\mathcal P}$ interchangeably and use $t\in{\mathcal P}$ to denote a term on the unique set of paths ${\mathcal P}$. The prior probability of a particular class ${\mathcal P}$ is $\pi_{{\mathcal P}}$, and is just the proportion of all objects within the class ${\mathcal P}$. If, within a data set we have a total of $|{\mathcal O}|$ objects and $|{\mathcal O}_{\mathcal P}|$ objects from class ${\mathcal P}$, then we can construct a simple estimator, $\hat{\pi}_{{\mathcal P}}$, of $\pi_{{\mathcal P}}$ via,

\begin{equation}
\hat{\pi}_{{\mathcal P}}\;=\; \frac{|{\mathcal O}_{{\mathcal P}}|}{|{\mathcal O}|}\;\; .
\label{eq:T.1}
\end{equation}

\noindent Clearly, we expect $\hat{\pi}_{\mathcal P}\rightarrow \pi_{\mathcal P}$ as $|{\mathcal O}|\rightarrow \infty$. We then use ${\mathcal N}_{t}$ to denote the expected number of classes within the data set to which a term $t$ belongs, {\it i.e.}, ${\mathcal N}_{t}$ is simply the number of leaves which are descendants of term $t$ in the ontology, weighted by the prior $\pi_{{\mathcal P}}$. Formally we define,

\begin{equation}
{\mathcal N}_{t}\;=\;\sum_{{\mathcal P}\ni t} \pi_{\mathcal P}\;\;,
\label{eq:T.2}
\end{equation}

\noindent and denote by $\widehat{{\mathcal N}}_{t}$ its equivalent defined using the estimators $\hat{\pi}_{{\mathcal P}}$, {\it i.e.},

\begin{equation}
\widehat{{\mathcal N}}_{t}\;=\;\sum_{{\mathcal P}\ni t} \hat{\pi}_{\mathcal P}\;\;.
\label{eq:T.2b}
\end{equation}

\noindent The value of ${\mathcal N}_{t}$ is essentially the probability, $P(t)$, that term $t$ is relevant to an object, irrespective of any further details of the object. Thus ${\mathcal N}_{t}$ gives a probabilistic measure of the specificity of term $t$ that reflects the structure of the ontology, and expert knowledge and information encoded within it. We therefore consider ${\mathcal N}_{t}$ to measure the intrinsic information content of term $t$. Specifically, we define the information content of term $t$ as $-\log P(t) = -\log {\mathcal N}_{t}$ \cite{CoverThomas1991}.

An object belonging to the class represented by a leaf ${\mathcal P}$ can be annotated with any of the terms $t$  belonging to  ${\mathcal P}$. We use $n_{ot} \in \{0,1\}$ to denote whether term $t$ is used to annotate object $o$. The discrete variables $n_{ot}$ are lattice gas occupancy numbers (or equivalently Ising spins). Consequently, we formulate the probability of a particular choice of annotations as a lattice-gas model. An example lattice-gas configuration corresponding to an example annotation is shown by the schematic on the right-hand side of Fig.\ref{fig:OntologySchematics}. The lattice-gas Hamiltonian is relatively simple and is given by,

\begin{equation}
H \;=\; \sum_{o,t} n_{ot}( v_{ot} - \mu )\;\; .
\label{eq:T.3}
\end{equation}

\noindent Here $v_{ot}$ represents the local field acting upon a particle at term $t$ for object $o$, and determines how likely it is that term $t$ will be selected when annotating object $o$. Ideally no mis-annotation occurs, {\it i.e.} only terms appropriate to each object are selected, and there would be the same probability of selecting term $t$ provided the paths ${\mathcal P}$ to which the object belongs contains $t$. Thus, we set,

\begin{equation}
v_{ot} \; = \; \left \{
\begin{array}{cc}
v_{t} & t \in {\mathcal P}\;{\rm and}\; o \in {\mathcal P}\;\;, \\
h & {\rm otherwise}\;\; .
\end{array}
\right .
\label{eq:T.4}
\end{equation}

\noindent The value of $h$ determines the global background level of mis-annotation that may occur. It is possible to consider a more structured form for $v_{ot}$, {\it e.g.}, to allow for greater likelihood of mis-annotation when a term $t$ is close to, but not in, the paths ${\mathcal P}$ to which the object $o$ belongs. However, for the remainder of this paper we consider the more simplified form for $v_{ot}$ given in Eq.(\ref{eq:T.4}) above and also only consider the scenario where no mis-annotation occurs. Consequently, we set $h = \infty$. Therefore, for each object, particular terms are forbidden if the term is not on any of the paths from the root node to the leaf node associated with the object. This is also illustrated in Fig.\ref{fig:OntologySchematics}. In practice, when analysing real data sets, it may be required to consider a finite value of $h$, to capture the mis-annotations that will inevitably occur.

Within Eq.(\ref{eq:T.3}) we have also included a chemical potential, $\mu$, which acts as a global field acting on all lattice cells. Equivalently, the fugacity $z=\exp(\beta\mu)$ is essentially the global prior probability that a term will be selected by an annotator, and controls or limits the average number, $\bar{n}$, of terms per object selected by an annotator. We have used $\beta = 1/T$ to denote the inverse temperature. The temperature $T$ does not at this stage have an explicit physical interpretation, other than to represent a parameter which controls the expectation value of the Hamiltonian in Eq.(\ref{eq:T.3}). However, in Section \ref{sec:MaxEntField} we propose an intrinsic form for the external field $v_{t}$ in terms of the information content of the term $t$. With this intrinsic form for $v_{t}$ the temperature $T$ then controls the average information retrieved from the ontology when an object is annotated. Clearly the temperature $T$ will also determine the level of variation seen in annotations from otherwise identical annotators when annotating the same set of objects.

Since we only have non-interacting particles, the probability $p_{t}$, of term $t$ being used to annotate an object for which it is valid, is easily calculated as,
\begin{equation}
p_{t}\;=\; \langle n_{ot}\rangle \;=\;\frac{e^{-\beta( v_{t} - \mu )}}{1 + e^{-\beta( v_{t} - \mu )}}\;\; .
\label{eq:T.5}
\end{equation}

\noindent Here $\langle n_{ot}\rangle$ denotes the expectation (over annotations) of $n_{ot}$. Likewise, with the simplification in Eq.(\ref{eq:T.4}) for $v_{ot}$ the partition function is straight-forward to evaluate, and for a particular set of objects is given by,
\begin{equation}
\log Z \;=\; |{\mathcal O}| \sum_{t}\widehat{{\mathcal N}}_{t}\log \left ( 1+e^{-\beta (v_{t} - \mu)}\right )\;\; .
\label{eq:T.7}
\end{equation}

\noindent Consequently, the expected value of $\log Z$, averaged over all possible data sets, ${\mathcal O}$, of size $|{\mathcal O}|$, is,

\begin{equation}
\langle \log Z \rangle _{\mathcal O} \;=\; |{\mathcal O}| \sum_{t}{\mathcal N}_{t}\log \left ( 1+e^{-\beta (v_{t} - \mu)}\right ) \;\; .
\label{eq:T.7b}
\end{equation}

\noindent Likewise, the expected frequency, $C_{t}$, of term $t$ within the annotation data set is given by,

\begin{equation}
C_{t}\;=\; \sum_{o\ni {\mathcal P}\ni t} \langle n_{ot} \rangle \;=\; -\beta^{-1}\frac{\partial \log Z}{\partial v_{t}} \;=\; |{\mathcal O}| \frac{e^{-\beta (v_{t}-\mu)}}{1+e^{-\beta ( v_{t} - \mu )}} \widehat{{\mathcal N}}_{t}\;=\; |{\mathcal O}| p_{t}\widehat{{\mathcal N}}_{t}\;\; ,
\label{eq:T.8}
\end{equation}

\noindent with the expectation over all data sets of size $|{\mathcal O}|$ being,

\begin{equation}
\langle C_{t}\rangle _{\mathcal O} \;=\; |{\mathcal O}| \frac{e^{-\beta (v_{t}-\mu)}}{1+e^{-\beta ( v_{t} - \mu )}}{\mathcal N}_{t}\;=\; |{\mathcal O}| p_{t}{\mathcal N}_{t}\;\; .
\label{eq:T.8b}
\end{equation}

\noindent In all of the expressions above for the partition function $Z$ we do not consider objects within the same class ${\mathcal P}$ to be interchangeable. It is likely that objects within the class ${\mathcal P}$ are unique and distinguishable, and simply possess a common characteristic rather than being replicates of each other.

Finally, we note that for any real data set of term frequencies $f_{t}$, Eq.(\ref{eq:T.8b}) provides a simple means of constructing an estimate of the field strength operating on node $t$, by equating the expectation value for $\langle C_{t}\rangle _{\mathcal O}$ given in Eq.(\ref{eq:T.8b}) with $f_{t}$. This gives us an estimate $\hat{v}_{t}$ for $v_{t}$ given by,

\begin{equation}
\hat{v}_{t}\;=\;\beta^{-1}\left [ \log z \;+\; \log \left (\frac{|{\mathcal O}|{\mathcal N}_{t}}{f_{t}} - 1 \right ) \right ]\;\; .
\label{eq:T.8b1}
\end{equation}

\section{Replicating the patterns of real annotation data sets \label{sec:realworldPatterns}}

\subsection{The relation between ${\mathcal N}_{t}$ and leaf descendants \label{sec:LeafDesc}}
Clearly, a first test of the lattice-gas model of the annotation process is whether it can replicate or explain the broad patterns seen in real annotation data sets, such as those seen by Kalankesh {\it et al.} \cite{Kalankesh2012}. From Eq.(\ref{eq:T.8b}) we can see that $\langle C_{t}\rangle _{\mathcal O}$ depends upon two factors; ${\mathcal N}_{t}$ and $p_{t}$. Due to the typically hierarchical nature of the ontology topology we would expect ${\mathcal N}_{t}$ to have a broad distribution of values, and potentially to dominate the distribution of $\langle C_{t}\rangle _{\mathcal O}$. It is transparent that for class probabilities that are uniform across all possible classes then the value of ${\mathcal N}_{t}$ is simply proportional to the number of leaf descendants, $LD_{t}$, that can be reached from term $t$. For small deviations away from a uniform class probabilities we would still expect ${\mathcal N}_{t}$ and $LD_{t}$ to be approximately proportional. Larger deviations away from uniform class probabilities effectively represent a pruning of the DAG into a smaller topology, with leaf nodes that correspond to small class probabilities being effectively eliminated. Consequently, for any DAG topology we still consider $LD_{t}$ to give useful insight into the expected term frequency $\langle C_{t}\rangle _{\mathcal O}$, and in the next section we assess the likely distribution for $LD_{t}$ for tree-like ontologies.

\subsection{The distribution of the leaf descendant count $LD_{t}$.\label{sec:LeafDescDist}}
For irregular-trees precise evaluation of $LD_{t}$ given local information about the node $t$ can be framed in terms of Galton-Watson processes \cite{GaltonWatson1875, Kendall1966}, though there are few usefully applicable closed-form results available to us. In contrast, single parent regular trees are easier to study. For a single-parent regular tree with $b$ children per node we can label the layers of the tree from $l=0$ (at the root) to $l=L$ (at the leaves). The total number of nodes in the tree is $(b^{L+1}-1)/(b-1)$. The number of nodes in layer $l$ is simply $b^{l}$, whilst the number of leaf descendants for each node in layer $l$ is $b^{L-l}$, {\it i.e.}, within each layer there is a simple reciprocal relationship between node counts and leaf descendant counts, suggesting a Zipf's behaviour for the distribution of $LD_{t}$. More formally, for node $t$ in layer $l$, with leaf descendant count $LD_{t}=b^{L-l}$, it is then a simple matter to find,
\begin{eqnarray}
&&{\rm Fraction\;of\;nodes\;with\;leaf\;descendant\;count} \geq LD_{t}\;=\;\frac{b-1}{b^{L+1}-1}\sum_{k=0}^{l}b^{k} \nonumber \\
&=& \frac{b^{l+1}-1}{b^{L+1}-1}\;=\; \frac{b^{L+1}}{b^{L+1}-1}\left [\frac{1}{LD_{t}}\;-\;\frac{1}{b^{L+1}}\right ]\;\simeq\;\frac{1}{LD_{t}}\;\;,\;{\rm as}\;L\rightarrow\infty\;.
\label{eq:T.8b1_2}
\end{eqnarray}

\begin{figure*}
\begin{center}
\scalebox{0.23}{%
\includegraphics*[bb = 0 0 792 612]{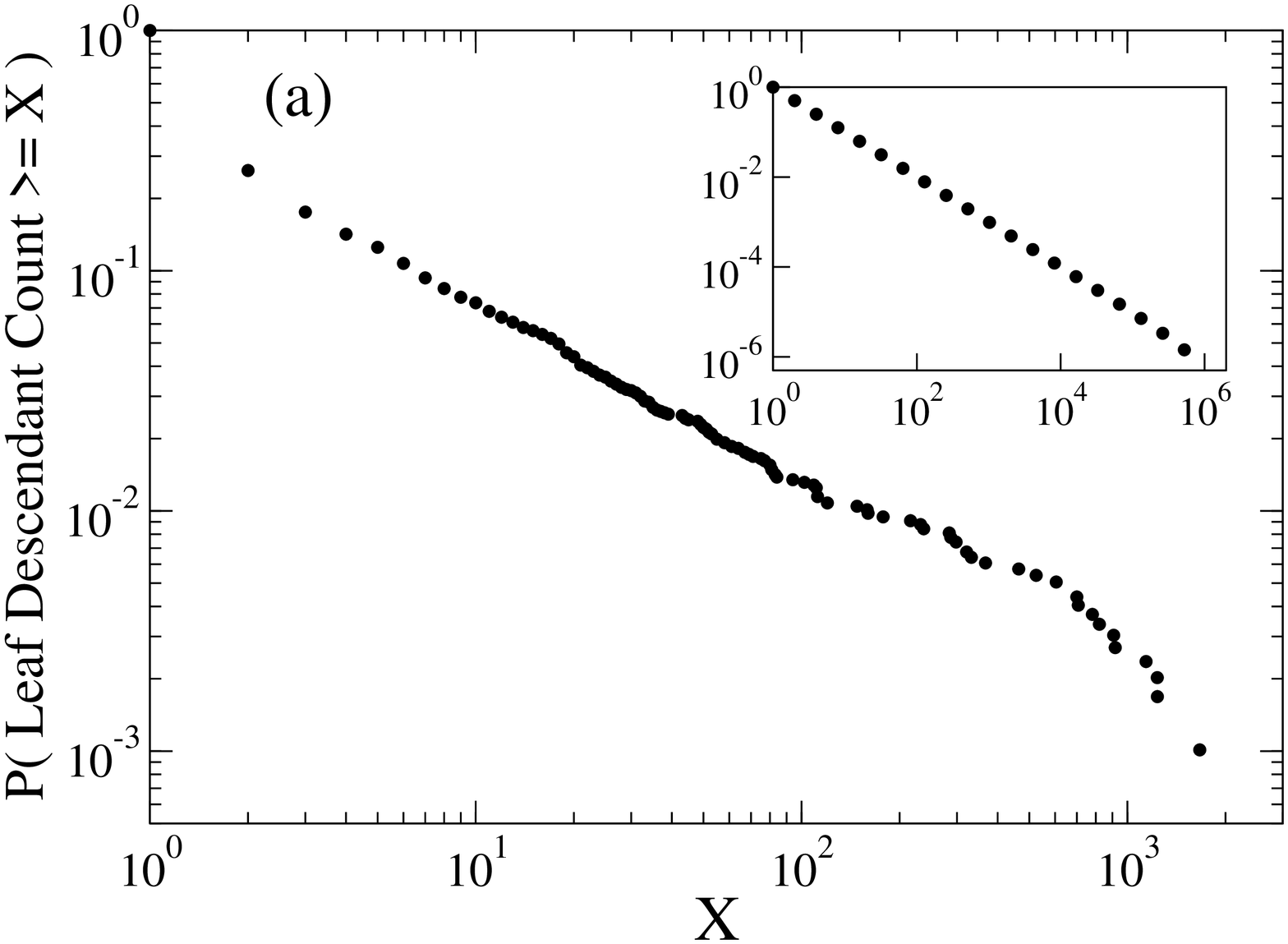}
\includegraphics*[bb = 0 0 792 612]{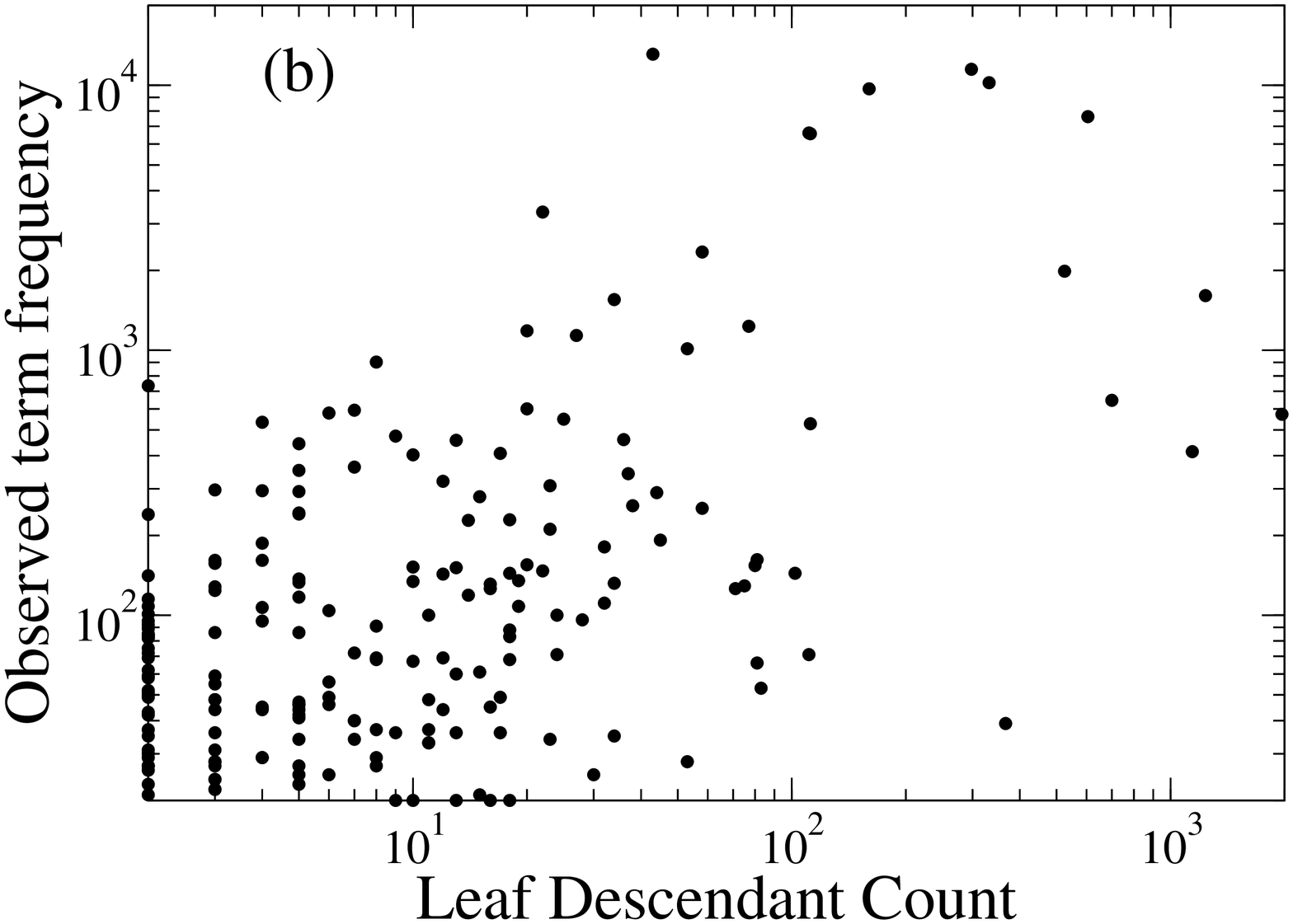}}
\end{center}
\caption{Plots of the distributions of leaf descendant counts for a real-world ontology and a regular tree.
Plot (a) shows a  plot of the log of the cumulative probability of $LD_{t}$ for all nodes with children within the GO cellular component ontology. The inset within plot (a) shows a plot of the log of the cumulative probability of $LD_{t}$ for all nodes with children within a bifurcating ($b=2$) regular tree of depth $L=20$. Plot (b) shows a log-log plot of observed term frequencies, $f_{t}$, against number of leaf descendants, $LD_{t}$, for the real data set shown in Fig.\ref{fig:KalankeshExample}.}
\label{fig:GOCC_Intrinsic}
\end{figure*}

\noindent Consequently, for large regular trees we will have $P(LD_{t} \ge X)\;\simeq X^{-1}$, {\it i.e.}, a Zipf's law form. Overall, Eq.(\ref{eq:T.8b1_2}) suggests the origin of the Zipf's law behaviour observed by Kalankesh {\it et al.} \cite{Kalankesh2012} may be, in part, a consequence of the Zipf's law like behaviour of the distribution of $LD_{t}$. However, whilst the structure of a regular tree will lead, on average, to a classical Zipf's law like behaviour, the degeneracy in leaf descendant count for nodes within the same layer leads to a very evenly spaced behaviour in any plot of the cumulative probability. This can be seen in the inset of Figure \ref{fig:GOCC_Intrinsic}a which shows the cumulative probability distribution (on a logarithmic scale) of leaf descendant counts for a regular tree with $b=2$ and depth $L=20$.

For a real ontology, where we will have significant variation in the number of children and parents at the scale of individual nodes, we would expect to see more continuous cumulative probability plots for $LD_{t}$. We have confirmed with further simulations of irregular tree-like ontologies (results not shown) that, typically, continuous Zipf's law like behaviour in $LD_{t}$ results when one has local variation in the node characteristics, {\it e.g.}, variation in the average number of children per node or in the probability that a node has children. Similarly, the main part of Fig.\ref{fig:GOCC_Intrinsic}a shows the cumulative probability distribution plot for the leaf descendant count of the cellular component GO ontology. This is the ontology used for annotating the data set shown in Fig.\ref{fig:KalankeshExample}. The intrinsic power-law like behaviour is clearly evident, and so it perhaps unsurprising that we should observe power-law like behaviour in annotations based upon this ontology. As postulated, and in contrast to the distribution shown for the regular tree, this real-world ontology shows a more continuous spread of leaf descendant counts. However, as with the regular tree, the effective exponent of the approximate power-law form for the cellular component GO ontology is close to -1.

\subsection{The external field $v_{t}$: intrinsic and specific components}
We have argued that term frequencies in real annotation data sets will be broadly determined by $LD_{t}$. Figure \ref{fig:GOCC_Intrinsic}b shows a log-log plot for the observed term frequencies, $f_{t}$, of the data set shown in Fig.\ref{fig:KalankeshExample}, plotted against the number of leaf descendants within the ontology from which the annotation terms were drawn. The broad correspondence between observed log of the term frequencies and the log of the number of leaf descendants is clear, and is statistically significant (estimate of Pearson correlation $\hat{\rho} = 0.454,\; 95\% CI = [0.408, 0.498]$, $p < 10^{-8}$ for null hypothesis of $\rho = 0$).

However, we would not expect an exact correspondence with the real observed term frequencies, $f_{t}$, on the basis of ${\mathcal N}_{t}$ alone. The precise value of $\langle C_{t}\rangle _{\mathcal O}$ is determined by two factors; ${\mathcal N}_{t}$ and $p_{t}$, with $p_{t}$ determined by $v_{t}$. We might expect $v_{t}$ itself to have an intrinsic component determined by the ontology topology, but the scatter in Fig.\ref{fig:GOCC_Intrinsic}b suggests that we should decompose $v_{t}$ into an intrinsic contribution, $v^{(0)}_{t}$, and a specific residual contribution, $\Delta v_{t}$. That is we write,

\begin{equation}
v_{t} \;=\; v^{(0)}_{t}\;+\; \Delta v_{t}
\label{eq:T.8b2}
\end{equation}

\noindent Once a form for $v^{(0)}_{t}$ has been specified and an estimate for $v_{t}$ has been obtained from observed term frequencies $f_{t}$ via Eq.(\ref{eq:T.8b1}), then an estimate for $\Delta v_{t}$ can be obtained. Overall, the values of $\Delta v_{t}$ reveal any biases towards particular terms beyond that expected on the basis of the ontology topology alone. Significant values of $\Delta v_{t}$ can potentially indicate regions of the ontology which are not matching the requirements of the annotators, and where the ontology can potentially be improved or needs modification. Equally, having a form for $v^{(0)}_{t}$ allows us to extend the theoretical statistical mechanical formalism by providing an explicit form for $p_{t}$ in the absence of any bias in $v_{t}$. Therefore, in the remainder of the paper we derive an appropriate form for $v^{(0)}_{t}$ and use it to study the consequences of increasing ontology complexity.

\section{Determination of the intrinsic field strength $v^{(0)}_{t}$ \label{sec:MaxEntField} }
To proceed we note that extracting or transferring information from the ontology to an object during the annotation process requires effort. Selecting more specific terms, far from the root, requires greater information about the object to be known by, or inferred by the annotator, and so we equate effort expended by an annotator with the intrinsic information content of the terms selected. A natural choice for $v^{(0)}_{t}$ would therefore be to set $v^{(0)}_{t}$ proportional to the information content of term $t$, {\it i.e.}, set $v^{(0)}_{t} \propto -\log {\mathcal N}_{t}$. This choice corresponds to the maximum entropy solution for the term probabilities $\{p_{t}\}_{t}$ under the constraint of capturing a given amount of information from the ontology using a fixed expected number of terms. To see this we first calculate the entropy, $S[\{ p_{t}\}]$ for for a given set of term probabilities, $\{p_{t}\}$. From Eq.(\ref{eq:T.5}) and Eq.(\ref{eq:T.7}) the entropy $S[\{ p_{t}\}]$ is obtained as,

\begin{eqnarray}
S[\{ p_{t}\}] & = & \langle \log Z\rangle_{\mathcal O}[\{v^{(0)}_{t}\}_{t}] \;+\;\beta\sum_{o,t}\langle \langle n_{ot}\rangle\rangle_{\mathcal O} ( v^{(0)}_{t} - \mu ) \nonumber \\
& = & -|{\mathcal O}| \sum_{t}{\mathcal N}_{t}\left [ p_{t}\log p_{t}\;+\; (1-p_{t})\log(1- p_{t}) \right ] \;\; .
\label{eq:T.9}
\end{eqnarray}

\noindent In calculating the entropy in Eq.(\ref{eq:T.9}) we have clearly averaged over data sets of size $|{\mathcal O}|$, as we wish to eliminate the effects of data set to data set variation and focus solely on the effect of data set size. If we wish to obtain the set of term probabilities $\{p_{t}\}$ which maximize the entropy subject to capturing a given amount of the stored information in the ontology and a given expected total number of selected terms, then we maximize,

\begin{eqnarray}
&& -\lambda\sum_{o,t}\langle\langle n_{ot}\rangle \log {\mathcal N}_{t}\rangle_{\mathcal O} \;+\; \phi\sum_{o,t}\langle\langle n_{ot}\rangle\rangle_{\mathcal O} \;+\;S\left [ \{p_{t}\}\right ] \nonumber \\
& = & -|{\mathcal O}|\sum_{t}{\mathcal N}_{t}p_{t}(\lambda\log {\mathcal N}_{t} - \phi)\; - \; |{\mathcal O}| \sum_{t}{\mathcal N}_{t}\left ( p_{t}\log p_{t}\;+\; (1-p_{t})\log(1- p_{t}) \right )\;\; . \nonumber \\
&&
\label{eq:T.10}
\end{eqnarray}

\noindent Here $\lambda$ and $\phi$ are Lagrange multipliers. The maximum entropy term probabilities, $p^{*}_{t}$, are simply the set of term probabilities that maximize the expression in Eq.(\ref{eq:T.10}), and so we find,

\begin{equation}
p^{*}_{t}\;=\; \frac{e^{-(\lambda\log {\mathcal N}_{t} -\phi)}}{1 + e^{-( \lambda\log {\mathcal N}_{t}-\phi)}}\;\; .
\label{eq:T.11}
\end{equation}

\noindent Comparing Eq.(\ref{eq:T.5}) and Eq.(\ref{eq:T.11}) we immediately see that this maximum entropy approach corresponds to setting,

\begin{equation}
\beta v^{(0)}_{t}\;=\; \lambda\log {\mathcal N}_{t}\;\;\;,\;\;\;\beta \mu\;=\;\phi\;\; .
\label{eq:T.11b-1}
\end{equation}

\noindent With $\beta >0$ we impose the relation (without loss of generality) $\beta = |\lambda|$, giving $v^{(0)}_{t} = -\log {\mathcal N}_{t}$ when $\lambda < 0$ and $v^{(0)}_{t} = \log {\mathcal N}_{t}$ when $\lambda > 0$. For $\lambda < 0$ terms with lower information content (larger ${\mathcal N}_{t}$), that are closer to the root, will be preferentially selected during annotation, whilst for $\lambda > 0$ terms with smaller values of ${\mathcal N}_{t}$, that are closer to the leaves of the ontology are preferentially selected.

Substituting the expression for $p^{*}_{t}$ from Eq.(\ref{eq:T.11}) into Eq.(\ref{eq:T.8}) we get the corresponding maximum entropy expected term frequencies as,
\begin{equation}
\langle C^{*}_{t}\rangle_{\mathcal O} \;=\;|{\mathcal O}|\frac{z{\mathcal N}_{t}^{1-\lambda}}{1 + z{\mathcal N}_{t}^{-\lambda}}\;\; .
\label{eq:T.12}
\end{equation}

\noindent Likewise, for the optimal term probabilities $\{p^{*}_{t}\}$, the free energy takes a value $F^{*}$ given by,
\begin{equation}
\beta F^{*} \; = \; \beta F[\{ p^{*}_{t}\}_{t}, \{\beta v_{t}=v^{(0)}_{t}\}_{t}]\;=\; -|{\mathcal O}|\sum_{t}{\mathcal N}_{t}\log ( 1 + z{\mathcal N}_{t}^{-\lambda} ) \;\; .
\label{eq:T.14}
\end{equation}

Having obtained a suitable form for $v^{(0)}_{t}$ it is a simple matter to determine whether, for any real data set, there is a non-zero specific field component $\Delta v_{t}$. This can be done by testing if the observed frequency $f_{t}$ is statistically significantly different from $\langle C^{*}_{t}\rangle_{\mathcal O}$ given by Eq.(\ref{eq:T.12}). If $f_{t}$ is significantly different to $\langle C^{*}_{t}\rangle_{\mathcal O}$, then we can then combine Eq.(\ref{eq:T.8b1}), Eq.(\ref{eq:T.8b2}), and Eq.(\ref{eq:T.11b-1}) to construct an estimate $\widehat{\Delta v}_{t} = \hat{v}_{t}\;-\;v^{(0)}_{t}$ for $\Delta v_{t}$. Thus, we have,

\begin{equation}
\widehat{\Delta v}_{t} \; = \; \left \{
\begin{array}{cc}
\frac{1}{|\lambda|}\left [ \log \left ( z{\mathcal N}^{-\lambda}_{t} \right ) \;+\; \log \left ( \frac{|{\mathcal O}|{\mathcal N}_{t}}{f_{t}} - 1 \right ) \right ] & {\rm if}\; f_{t}-\langle C^{*}_{t}\rangle_{\mathcal O}\;\; {\rm significant} \\
0 & {\rm otherwise}\;\; .
\end{array}
\right .
\end{equation}

Realistically, for a large ontology we would only expect a small proportion of all the terms to be used when annotating an object, and so from Eq.(\ref{eq:T.12}) we would expect $\langle C^{*}_{t}\rangle_{\mathcal O} / |{\mathcal O}| \ll 1$. This suggests that $z{\mathcal N}_{t}\ll 1$ and that in general $\langle C^{*}_{t}\rangle_{\mathcal O} \sim {\mathcal N}_{t}^{1-\lambda}$. This gives us a simple mechanism for estimating appropriate values of $\lambda$ for real data sets. This also tells us that, with ${\mathcal N}_{t}$ from tree-like ontologies expected to display a Zipf's law behaviour with an exponent close to -1, then we expect to see a Zipf's law like behaviour in the observed term frequencies $f_{t}$ with exponent $-(1-\lambda)$. As the exponents for $f_{t}$ observed by Kalankesh {\it et al.} \cite{Kalankesh2012} are close to -1, this suggests that the effective values of $\lambda$ for real data sets are small in magnitude and potentially either side of $\lambda=0$.

Finally, we note that as $\lambda = 0$ represents the boundary at which terms closer to the root are preferentially selected, an effective value of $\lambda < 0$ within a real data set may suggest that annotators do not have a strong desire to make use of the additional nodes provided by larger complex ontologies. Within our statistical mechanical framework we are able to explore whether there is a `optimal' or preferred size to the ontology that annotators wish to use, and also how this `optimal' size might scale with the number of objects being annotated. This we do in the next section.

\section{Statistical mechanics of an ensemble of ontologies \label{sec:Metric}}
Having set up a statistical mechanical model of the annotation process, we are in a position to perform more theoretical analyses and experiments that, whilst not being realisable, still provide valuable insight into the performance of annotators and the underlying ontology being used. For example, a key practical question we wish to address is to determine whether a real-world ontology is fit for purpose, or whether the ontology needs shrinking or expanding. To answer this question in real terms would require providing annotators with a range of ontology graphs and observing which ontology is used most frequently. Such an experiment is not readily performed in real terms. However, within the statistical mechanical framework we can address this question by extending the previous ensemble to one consisting of a large collection of annotators who annotate the same set of objects, but who can select the ontology to be used from a range of ontologies of differing complexity. Consequently, we will see variation in both the choice of topology ${\mathcal H}$ being used, and the annotation configuration $\{ n_{ot}\}$ being selected. Considering this ensemble allows us to easily determine the most preferred, and hence most appropriate, ontology. This optimal ontology will depend upon a number of parameters, such as the number of objects being annotated, $|{\mathcal O}|$, and annotator characteristics such as the average number of terms used per object and the average effort/information the annotator is willing to expend per object. The latter two characteristics are controlled by the parameters $z$ and $\lambda$. Therefore, the remainder of Section \ref{sec:Metric} is focused on elucidating how the preferred, or optimal, ontology varies with the parameters $|{\mathcal O}|, z$ and $\lambda$. To determine the variation with respect to $|{\mathcal O}|, z$ and $\lambda$ requires breaking the analysis down into a number of smaller, but still quite involved, steps. Firstly, in Section \ref{sec:complexityQuant} we construct a suitable statistical mechanical potential, $\beta \Omega^{*}$, from which the optimal ontology topology ${\mathcal H}$ can be determined. In section \ref{sec:regTrees} we restrict ${\mathcal H}$ to the class of regular trees to make analysis of $\beta \Omega^{*}$ tractable. In Section \ref{sec:asympAnalysis} we further simplify the analysis by developing a closed form approximation to $\beta \Omega^{*}$ for regular trees, and examining the asymptotic behaviour of the approximation as the number of objects $|{\mathcal O}|$ increases. In Section \ref{sec:simulation} we confirm the results of the asymptotic analysis on regular trees using numerical simulations. Finally, in Section \ref{sec:lambdaTrend} we use the insights gained from the analysis of $\beta \Omega^{*}$ to understand what are the most likely observed values for the parameter $\lambda$.

\subsection{Quantifying ontology complexity and determination of the optimal ontology \label{sec:complexityQuant} }
To extend the ensemble to one that consists of ontologies of varying complexity we must first introduce a measure to quantifying the intrinsic complexity of each ontology. Continuing the information theoretic approach, we measure the complexity of an ontology by its total intrinsic information content $-\sum_{t}\log {\mathcal N}_{t}$, and so introduce an additional control variable, $\omega > 0$, to set the average intrinsic information content of a ontology node within this ensemble. The partition function for this ensemble becomes,
\begin{equation}
{\mathcal Z}\;=\; \sum_{{\mathcal H}} \exp \left ( {\omega \sum_{t}\log {\mathcal N}_{t}}\right ) Z({\mathcal H})\;\; ,
\end{equation}

\noindent where $Z({\mathcal H})$ is the partition function on a fixed DAG  and $\log Z({\mathcal H})$ is given by $-\beta F^{*}$ in Eq.(\ref{eq:T.14}). Larger values of $\omega$ will favour smaller values for the average intrinsic information per node, and so will favour smaller, lower complexity ontologies. Just as the partition function has been modified, then similarly, the appropriate potential for this ensemble is obtained by adding $-\omega \sum_{t}\log {\mathcal N}_{t}$ to the optimal value of the free energy, $\beta F^{*}$. This gives a potential,

\begin{equation}
\beta \Omega^{*} \;=\; -|{\mathcal O}|\sum_{t}{\mathcal N}_{t}\log ( 1 + z{\mathcal N}_{t}^{-\lambda} )\;-\;
\omega \sum_{t}\log {\mathcal N}_{t}\;\; .
\label{eq:M.2}
\end{equation}

\noindent The most frequently selected, or optimal, ontology is that which minimizes $\beta\Omega^{*}$. The term $-\omega \sum_{t}\log {\mathcal N}_{t}$ therefore serves an important function by regulating the ontology complexity, with more complex topologies being penalized more heavily.

We can see that the minimum value  $\beta\Omega^{*}(\{ {\mathcal N}_{t}\}_{t}, z, \omega)$ changes on varying either the topology ${\mathcal H}$ or the total number of objects, $|{\mathcal O}|$, to which the ontology applies. Consequently, we have the possibility of a change in optimal DAG topology as the number of objects, $|{\mathcal O}|$, being annotated increases. As previously stated, this possibility of a change in the optimal DAG topology is something we wish to investigate further, though to do so over a range of general DAG topologies is unlikely to be tractable. However, as we have already argued for an ontology to be effective it will have a tree-like topology. This then highlights the importance of understanding the behaviour of $\beta\Omega^{*}$ for tree-like ontologies. Even so, characterizing the behaviour of $\beta\Omega^{*}$ for an arbitrary tree-like ontology is still likely to be a difficult task. Therefore, to gain further insight into the behaviour of $\beta \Omega^{*}$ for tree-like ontologies we restrict our further analysis of $\beta \Omega^{*}$ to regular trees.

\subsection{Optimal ontology size for regular trees \label{sec:regTrees} }
For a regular tree with multi-furcating nodes, {\it i.e.}, each parent having $b$ children, we can label nodes according to which layer, $l$, the node is in. The root node is in layer $l=0$, whilst leaf nodes are in the final layer $L$.  Nodes in layer $l$ have $b^{L-l}$ leaf descendants and there are $b^{l}$ nodes in layer $l$. Consequently, the total number of leaves is $b^{L}$. For simplicity we will take the class distribution $\{\pi_{\mathcal P}\}_{\mathcal P}$ to be uniform across the leaf nodes, {\it i.e.}, $\pi_{\mathcal P} = b^{-L}$ for each of the classes corresponding to a leaf node. From these relations we can evaluate ${\mathcal N}_{t} = b^{-l}$,  for a term corresponding to a node in layer $l$. It is then a simple matter to re-write $\beta\Omega^{*}$ as,

\begin{equation}
\beta \Omega^{*}\;=\; -|{\mathcal O}| \sum_{l=0}^{L} \log \left ( 1 + zb^{l\lambda}\right ) \;+\; \omega \log b \sum_{l=0}^{L} lb^{l}\;\; .
\label{eq:R.1}
\end{equation}

\noindent The average number, $\bar{n}$, of terms used per annotated object is determined by the fugacity $z$ via the relation,

\begin{equation}
\bar{n}\;=\; -\frac{z}{|{\mathcal O}|}\frac{\partial}{\partial z} \beta \Omega^{*}\;=\; \sum_{l=0}^{L} \frac{zb^{l\lambda}}{1+zb^{l\lambda}}\;\; .
\label{eq:R.2}
\end{equation}

\noindent Thus, if we wish to attain a specified value $\bar{n}$ we simply solve Eq.(\ref{eq:R.2}) for the required value $z$. Similarly, the average information retrieved per object, $I$, is determined via the relation,

\begin{equation}
I\;=\; -\frac{1}{|{\mathcal O}|}\frac{\partial}{\partial \lambda} \beta \Omega^{*}\;=\;\log b\sum_{l=0}^{L} \frac{zlb^{l\lambda}}{1+zb^{l\lambda}}\;\; .
\label{eq:R.2b}
\end{equation}

The simple form for the $\beta\Omega^{*}$ in Eq.(\ref{eq:R.1}) also allows us to analytically determine the tree size $L$ that is optimal for annotating a given number of objects $|{\mathcal O}|$. The optimal tree size is that which minimizes $\beta\Omega^{*}$. Due to the discrete nature of $L$, any growth we observe in the optimal ontology size will occur via a series of transitions. Naively, we would expect the optimal ontology size, $L_{opt}$, to increase as the number of annotated objects is increased. That is, we would {\it a priori} expect $L_{opt}\rightarrow\infty$ as $|{\mathcal O}|\rightarrow\infty$. However, whether growth of $L_{opt}$ is possible or not may be affected by the particular value of $z$ or $\lambda$. Consequently, in the next section our analysis will focus upon the behaviour of $\beta\Omega^{*}$ as $|{\mathcal O}|$ and $L$ increase, in particular in the regime $|{\mathcal O}|, L\rightarrow\infty$, for different choices of $z$ and $\lambda$.

\subsection{Asymptotic behaviour of the optimal ontology size for regular trees \label{sec:asympAnalysis} }
The second term on the right-hand-side of Eq.(\ref{eq:R.1}) is easily evaluated as,
\begin{eqnarray}
\omega \log b \sum_{l=0}^{L} lb^{l} & = & \omega \frac{\log b}{(b-1)^{2}} \left [ Lb^{L+2} \;-\;(L+1)b^{L+1} + b \right ] \nonumber \\
& \simeq & \frac{\omega \log b}{(b-1)}Lb^{L+1}\;\;\;,\;{\rm as}\;L\rightarrow\infty\;\; .
\label{eq:R.3}
\end{eqnarray}

\noindent To evaluate the first term on the right-hand-side of Eq.(\ref{eq:R.1}) we define $a=b^{\lambda}$ and make use of the Euler-Maclaurin sum formula \cite{WhittakerWatson1990} to obtain (see \ref{sec:EulerMaclaurin} for details),

\begin{eqnarray}
\sum_{l=0}^{L} \log \left ( 1 + za^{l}\right ) & \simeq & \frac{1}{\log a}\left ( Li_{2}(-z)\;-\;Li_{2}(-za^{L})\right )
\;+ \; \frac{1}{2}\log \left( 1+za^{L}\right ) \nonumber \\
& + &\frac{1}{2}\log\left (1+z\right ) \; + \; \frac{\log a}{12}\left ( \frac{za^{L}}{1+za^{L}}\;-\;\frac{z}{1+z}\right )\;\; .
\label{eq:R.3e}
\end{eqnarray}

\noindent From the approximation in Eq.(\ref{eq:R.3e}) we have,
\begin{eqnarray}
\bar{n} \;=\;\frac{1}{\log a}\log \left ( \frac{1+za^{L}}{1+z}\right ) & + & \frac{1}{2}\frac{za^{L}}{1+za^{L}}\;+\;\frac{1}{2}\frac{z}{1+z}\nonumber \\
& + & \frac{\log a}{12}\left (\frac{za^{L}}{(1+za^{L})^{2}}\;-\;\frac{z}{(1+z)^{2}} \right )\;\; .
\label{eq:R.3f}
\end{eqnarray}

\noindent For convenience of later analysis we can also write Eq.(\ref{eq:R.3e}) as,
\begin{equation}
\sum_{l=0}^{L} \log \left ( 1 + za^{l}\right ) \simeq f_{1}(za^{L}) \;+\; f_{2}(z)\;\; .
\label{eq:R.3f2}
\end{equation}

\noindent with obvious definitions for the functions $f_{1}$ and $f_{2}$. With this more compact notation we can re-write Eq.(\ref{eq:R.3f}) as,
\begin{equation}
\bar{n}\;=\; za^{L}f^{'}_{1}(za^{L})\;+\; z f_{2}^{'}(z)\;\; .
\label{eq:R.3f3}
\end{equation}

The approximation developed in Eq.(\ref{eq:R.3e}) is extremely accurate for the values of $z,a$ and $L$ we are interested in (see \ref{sec:EulerMaclaurin} for details), and thereby allows us to accurately elucidate the growth behaviour of the optimal ontology size. To proceed, we consider the asymptotic behaviour of the approximation to $\beta\Omega^{*}$, as $|{\mathcal O}|$ and $L\rightarrow\infty$. There are two regimes potentially worth studying in the asymptotic limit $|{\mathcal O}|, L\rightarrow\infty$. The first regime is where $\bar{n}$ is fixed in size. In this regime the increasing number of terms available for annotating an object, as $L\rightarrow\infty$, are not made use of, {\it i.e.}, the increased information captured within the larger ontologies is effectively ignored. Consequently, we also find it instructive to consider a second regime where $z$ is fixed, resulting in $\bar{n}$ scaling linearly with $L$. Detailed analysis of the behaviour of $\beta\Omega^{*}$ under these two regimes is given below, for both $\lambda < 0$ and $\lambda > 0$.\newline

\noindent \underline{$\bm{\bar{n}}$ \bf{fixed as} $\bm {L\rightarrow\infty,\; \lambda < 0:}$}

\noindent For $\lambda < 0$ we have $a = b^{\lambda} < 1$ and so $a^{L}\rightarrow 0$ as $L\rightarrow\infty$. Therefore we can expand the right hand side of both Eq.(\ref{eq:R.3f2}) and Eq.(\ref{eq:R.3f3}) in powers of $a^{L}$. Doing so, we find,
\begin{equation}
\bar{n} \; = \; z_{0}f^{'}_{2}(z_{0})\;\; , \;\;z \; = \; z_{0} \;+\; O(a^{L})\;\; ,
\label{eq:R.4.3}
\end{equation}
\begin{equation}
-\frac{\beta F^{*}}{|{\mathcal O}|} \; = \; f_{1}(0)\;+\;f_{2}(z_{0}) \;+\; \frac{z^{2}_{0}f^{'}_{1}(0)f_{2}^{''}(z_{0})}{f_{2}^{'}(z_{0}) + z_{0}f_{2}^{''}(z_{0})}\,a^{L}\;+\; O(a^{2L})\;\; ,
\label{eq:R.4.2}
\end{equation}

\noindent and so the leading order contributions to $\beta\Omega^{*}$ take the form,
\begin{equation}
\beta\Omega^{*}\;\simeq\; {\rm Constant} \;-\;|{\mathcal O}|\frac{z^{2}_{0}f^{'}_{1}(0)f_{2}^{''}(z_{0})}{f_{2}^{'}(z_{0}) + z_{0}f_{2}^{''}(z_{0})}\,a^{L} \;+\;\frac{\omega \log b}{(b-1)}Lb^{L+1}\;\;\;{\rm as}\;L\rightarrow\infty\;\;.
\label{eq:R.5}
\end{equation}

\noindent The optimal tree depth $L_{opt}$ is determined by minimizing $\beta\Omega^{*}$ with respect to $L$. With the constant in Eq.(\ref{eq:R.5}) being independent of $L$ we find that, provided $\frac{z^{2}_{0}f^{'}_{1}(0)f_{2}^{''}(z_{0})}{f_{2}^{'}(z_{0}) + z_{0}f_{2}^{''}(z_{0})} < 0$, we have,
\begin{equation}
L_{opt} \;\sim\; \frac{\log |{\mathcal O}|}{(1-\lambda)\log b}\;\;\;,\;{\rm as}\;|{\mathcal O}|\rightarrow\infty\;\; .
\label{eq:R.5b}
\end{equation}

\noindent We show in \ref{sec:proof1} that for $ 0 > \log a > -6 -4\sqrt{6}\simeq -15.8$ we indeed have that  $\frac{z^{2}_{0}f^{'}_{1}(0)f_{2}^{''}(z_{0})}{f_{2}^{'}(z_{0}) + z_{0}f_{2}^{''}(z_{0})} < 0$. So in this regime for $\log a$ Eq.(\ref{eq:R.5b}) predicts growth of $L_{opt}$, via a series of transitions, as $|{\mathcal O}|$ is increased, with Eq.(\ref{eq:R.5b}) giving the global scaling relation between the optimal value $L_{opt}$ and $|{\mathcal O}|$. It is also worth noting that the scaling relation in Eq.(\ref{eq:R.5b}) predicts that the number of leaf nodes, $b^{L_{opt}}$, or equivalently the number of required classes, grows slower than the number of objects $|{\mathcal O}|$. The restriction to $\log a > -15.8$ is not an onerous one, as it is likely to be well outside the range of values for $a$ that we are interested in (see the end of \ref{sec:proof1} for a discussion of this point).\newline

\noindent \underline{$\bm{\bar{n}}$ \bf{fixed as} $\bm {L\rightarrow\infty,\; \lambda > 0:}$}

\noindent For $\lambda > 0$ we have $a=b^{\lambda} > 1$ and so from Eq.(\ref{eq:R.3f}) we can see that in order to maintain a fixed value of $\bar{n}$ as $L\rightarrow\infty$ we must have the scaling $z\sim a^{-L}$. Thus, in general we write $z=\hat{z}_{1}a^{-L} + \hat{z}_{2}a^{-2L} + O(a^{-3L})$. With $z \ge 0$ then we must have $\hat{z}_{1}\ge 0$ for this decomposition of $z$ to hold over an arbitrary range of $L$. Substituting this form for $z$ into Eq.(\ref{eq:R.3f}) we find,
\begin{equation}
\bar{n}\;=\; \hat{z}_{1}f^{'}_{1}(\hat{z}_{1})\;\;\;\;,\;\;\;\;\hat{z}_{2}\;=\;-\frac{\hat{z}_{1}f^{'}_{2}(0)}{f^{'}_{1}(\hat{z}_{1})+\hat{z}_{1}f^{''}_{1}(\hat{z}_{1})}
\;\; .
\label{eq:R.14}
\end{equation}

\noindent From this we find,
\begin{equation}
\beta \Omega^{*} \;\simeq\; -|{\mathcal O}|\left ( {\rm Constant} \;-\;a^{-L}\hat{z}_{1}\hat{z}_{2}f^{''}_{1}(\hat{z}_{1})\;+\; O(a^{-2L})\right )\;\;+\;\;\frac{\omega \log b}{(b-1)}Lb^{L+1}\;\; .
\label{eq:R.15}
\end{equation}

\noindent The constant in the expansion above does not depend upon $L$. In \ref{sec:proof2} we show that $\hat{z}_{1}\hat{z}_{2}f^{''}_{1}(\hat{z}_{1}) < 0$, leading to the immediate conclusion that, as $L\rightarrow\infty$, the leading order non-constant term in the asymptotic expansion of the free energy is positive, and so cannot counter-balance the increasing contribution from the complexity penalty term. That is, large tree depths $L$ will never be optimal (in the sense of producing a stationary value of $\beta\Omega^{*}$) for any value of $|{\mathcal O}|$, irrespective of the values of $\bar{n}$ and $a$. A natural corollary is that the optimal tree depth, $L_{opt}$, is then simply the smallest tree that will admit the required value of $\bar{n}$, {\it i.e.}, $L_{opt} = \bar{n} + 1$, though strictly speaking the leading order asymptotic analysis may not still be valid at such values of $L$. However, the fact that the asymptotic analysis suggests that using large trees is sub-optimal, irrespective of how many objects we wish to annotate, is surprising.\newline

\noindent \underline{$\bm{z}$ \bf{fixed as} $\bm {L\rightarrow\infty,\; \lambda < 0:}$}

\noindent With $a<1$ for $\lambda <0$ we have $za^{L}\rightarrow 0$ as $L\rightarrow\infty$, and so expanding $\beta\Omega^{*}$ in powers of $a^{L}$ gives,
\begin{equation}
\beta\Omega^{*}\;\simeq\; |{\mathcal O}| \left [ {\rm Constant}\;-\; za^{L}\left ( \frac{1}{\log a} + \frac{1}{2} +\frac{\log a}{12}\right )\right ] \;+\; \frac{\omega \log b}{(b-1)}Lb^{L+1}
\;\; .
\label{eq:R.5c}
\end{equation}

\noindent Again, the constant in the expansion of $\beta\Omega^{*}$ has no dependence upon $L$. As we have already shown in \ref{sec:proof1}, when $\log a < 0$ we have $\frac{1}{\log a}+\frac{1}{2} + \frac{\log a}{12} <0$. On setting the derivative (with respect to $L$) of Eq.(\ref{eq:R.5c}) to zero we obtain,
\begin{equation}
L_{opt} \;\sim\; \frac{\log |{\mathcal O}|}{(1-\lambda)\log b}\;\;\;,\;{\rm as}\;|{\mathcal O}|\rightarrow\infty\;\; .
\label{eq:R.5d}
\end{equation}

\noindent Again, we see that it is predicted that the number of classes grows slower than the number of objects $|{\mathcal O}|$.\newline

\noindent \underline{$\bm{z}$ \bf{fixed as} $\bm {L\rightarrow\infty,\; \lambda > 0:}$}

\noindent For $\lambda > 0$ we have $a=b^{\lambda} > 1$ and so for $z$ fixed we have $za^{L}\rightarrow\infty$ as $L\rightarrow\infty$. Then to obtain the asymptotic behaviour of $\beta\Omega^{*}$ we observe the following relations and representation for the dilogarithm $Li_{2}(x)$ \cite{Kirillov1995},
\begin{eqnarray}
&& Li_{2}(x) \;+\; Li_{2}(-x) \;=\; \frac{1}{2}Li_{2}(x^{2})\;\;\forall x\in \field{C}\;\;, \nonumber \\
&& Li_{2}(x)\;=\; \frac{\pi^{2}}{3}\;-\;\frac{1}{2}\left (\log x\right )^{2}\;-\;i\pi\log x \;-\;\sum_{k=1}^{\infty}\frac{1}{k^{2}x^{k}}\;\;\;\forall x \in \field{R}, x \geq 1\;\;.
\label{eq:R.6}
\end{eqnarray}

\noindent Utilising the results above for $Li_{2}(x)$ we finally arrive at the leading order behaviour of $\beta F^{*}$,
\begin{equation}
\frac{\beta F^{*}}{|{\mathcal O}|} \; = \; {\rm Constant}\;-\; \frac{L^{2}}{2}\log a\;-\;L\left ( \log z\;+\;\frac{1}{2}\log a \right )\; + \; O\left (a^{-L}\right )\;\; .
\label{eq:R.7}
\end{equation}

\noindent As before, the constant contains only terms that do not depend upon $L$. Setting $\left . \frac{\partial \beta\Omega^{*}}{\partial L}\right |_{L=L_{opt}} = 0$, we find the optimal tree depth $L_{opt}$ satisfies the leading order scaling relation,
\begin{equation}
L_{opt}\;\sim\; \frac{\log |{\mathcal O}|}{\log b}\;\;\;\; {\rm as}\; |{\mathcal O}|\rightarrow\infty\;\; .
\label{eq:R.11}
\end{equation}

\noindent In contrast to the previous scenarios, we now have $|{\mathcal O}|b^{-L_{opt}}\sim {\rm constant}$, {\it i.e.} the number of objects per class (leaf node) is approximately constant (or more correctly, only a weak function of $L_{opt}$). It is also interesting to observe that with $z$ fixed, the average number of terms used per object, $\bar{n}$, is given by,

\begin{equation}
\bar{n}\;=\; -\frac{z}{|{\mathcal O}|}\frac{\partial}{\partial z} \beta \Omega^{*} \; \simeq\; {\rm Constant} \;+\;L\;\;\;\;,\;\;{\rm as}\;L\rightarrow\infty\;\; .
\label{eq:R.8}
\end{equation}

\noindent Thus for fixed $z$ we find $\bar{n}$ scaling linearly with $L$. If we have a measurement of $\bar{n}$ at a particular tree depth $L_{0}$, then we can re-express Eq.(\ref{eq:R.8}) as,

\begin{equation}
\bar{n}(L)\;=\; \bar{n}(L_{0})\;+\;(L\;-\; L_{0})\;\; .
\label{eq:R.9}
\end{equation}

Summarizing the asymptotic behaviour across the four scenarios we have,
\begin{equation}
\begin{array}{ccccc}
L_{opt} \times \log b & \sim & \log |{\mathcal O}| / (1-\lambda)\;\;\; &,&\;\bar{n}\;{\rm fixed,}\;\lambda < 0\;\; ,  \\
L_{opt} \times \log b & \sim & {\rm Constant}\;\;\;&,&\;\bar{n}\;{\rm fixed,}\;\lambda > 0\;\; ,  \\
L_{opt} \times \log b & \sim & \log |{\mathcal O}| / (1-\lambda)\;\;\;&,&\;z\;{\rm fixed,}\;\lambda < 0\;\; , \\
L_{opt} \times \log b & \sim & \log |{\mathcal O}|\;\;\;&,&\;z\;{\rm fixed,}\;\lambda > 0 \;\; .
\end{array}
\label{eq:R.10}
\end{equation}

\noindent As estimates for $\lambda, \bar{n}$ (and hence $z$) can already be obtained, the scaling laws (along with their associated amplitudes) provide us with a potential mechanism for assessing whether a tree-like ontology is of optimal size for the given number, $|{\mathcal O}|$, of objects which it is being used to annotate. Thus, we can potentially assess whether an existing ontology should be expanded, or is overly complex for its current usage. Application of these scaling laws is clearly dependent upon their accuracy and the values of $\lambda$ we are likely to encounter for larger ontologies. These aspects we assess in the next two sections.

\subsection{Simulation validation of optimal ontology growth and scaling laws \label{sec:simulation} }

\begin{figure*}
\begin{center}
\scalebox{0.23}{%
\includegraphics*[bb = 0 0 792 612]{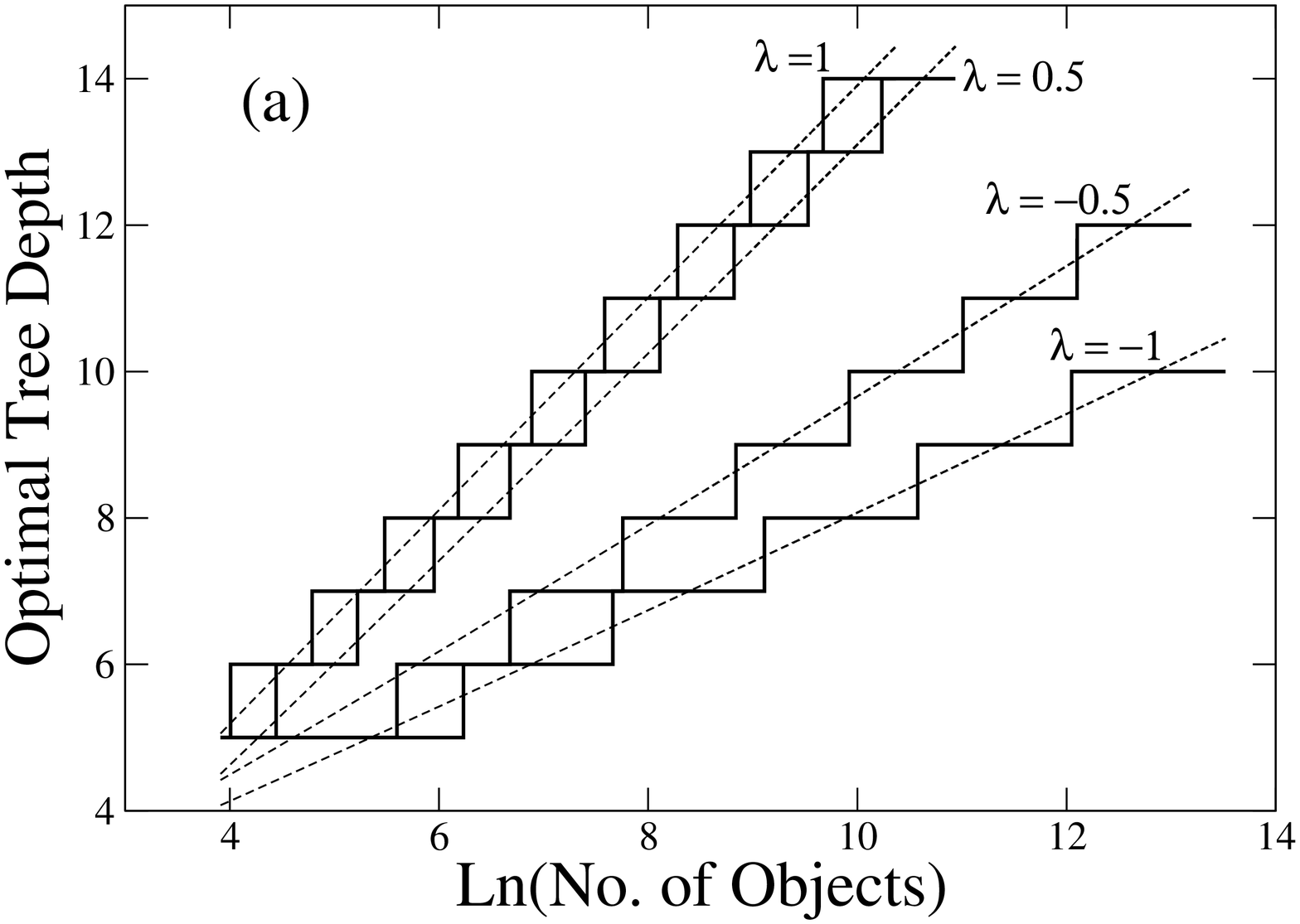}
\includegraphics*[bb = 0 0 792 612]{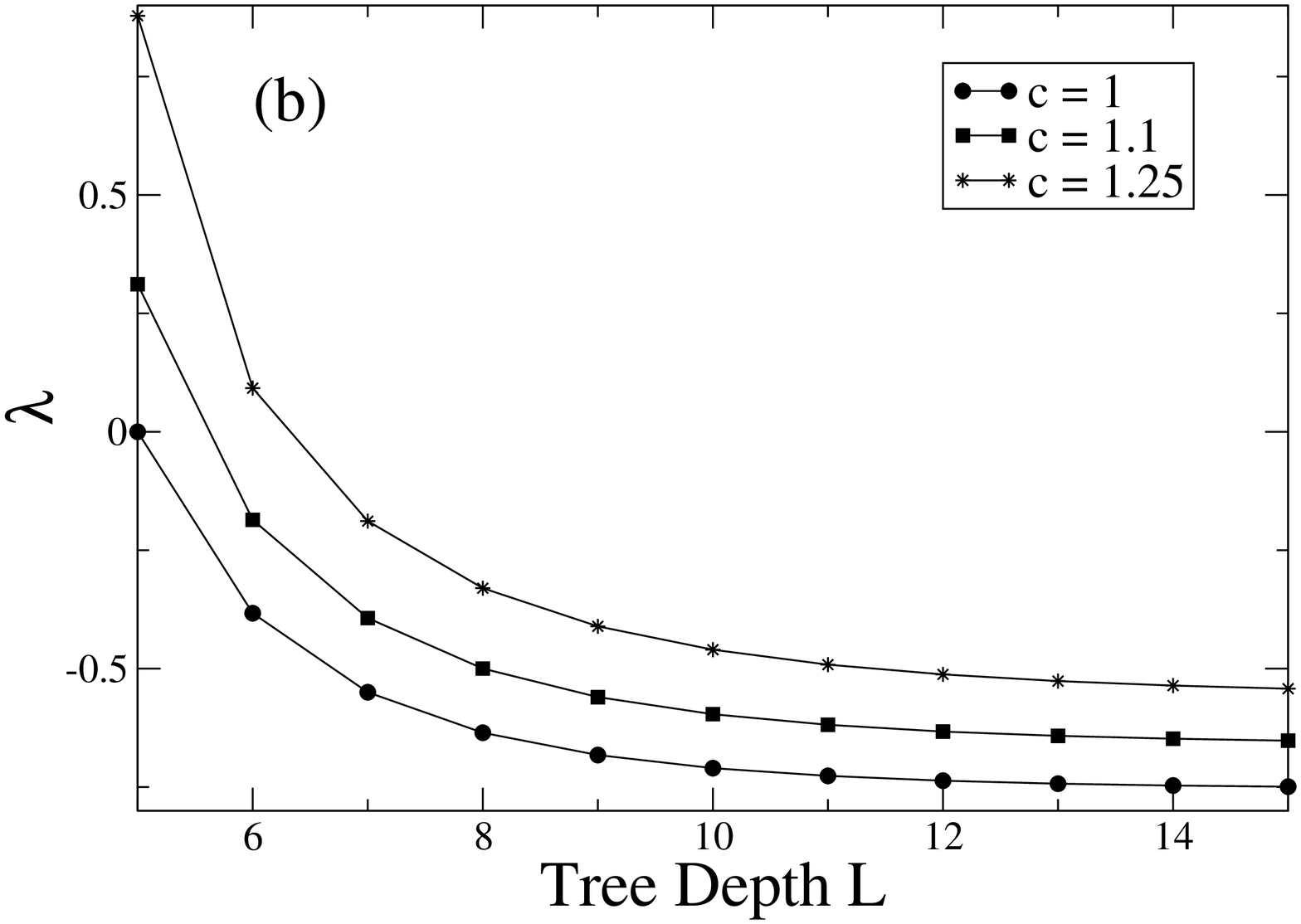}}
\end{center}
\caption{a) Plot of optimal tree depth $L_{opt}$ versus $\ln |{\mathcal O}|$ for various values of $\lambda$. The solid lines represent the optimal tree depth $L$ determined by finding the minimum of $\beta\Omega^{*}$ over integer values of $L$. The dashed lines represent the broad growth trend as predicted by minimizing the leading order asymptotic contributions to $\beta\Omega^{*}$. b) Plots of the required values of $\lambda$ as $L$ increases for a fixed given value of $I$ and $\bar{n}$.}
\label{fig:figR3}
\end{figure*}

The possibility of growth in the optimal tree depth $L_{opt}$ arises in three of the four distinct regimes considered above; namely at fixed $z$ for both $\lambda <0$ and $\lambda >0$, and also at fixed $\bar{n}$ for $\lambda < 0$. This is borne out by numerical calculations. In Figure \ref{fig:figR3}a we have shown the growth, with $|{\mathcal O}|$, in the tree depth $L_{opt}$. For fixed $\bar{n}$ with $\lambda < 0$ the fugacity $z$ tends to a finite non-zero value as $L\rightarrow\infty$, so essentially we can regard the fixed $\bar{n}, \lambda < 0$ regime as equivalent to the fixed $z$, $\lambda < 0$ regime. Therefore, we have performed the calculations at a number of different values for $\lambda$, but in all cases held the fugacity fixed to achieve a value of $\bar{n}=\frac{25}{7}$ at $L=5$. All calculations of $\beta\Omega^{*}$ have used the exact summation form for evaluation of $\beta F^{*}$. The dashed lines shown in Fig.\ref{fig:figR3}a correspond to the growth in $L_{opt}$ predicted by by minimizing the leading order asymptotic contributions to the integral approximation of $\beta\Omega^{*}$. The correspondence between the simulation results and the long term growth trend predicted by the asymptotic analysis is good, confirming the leading order scaling relations given in Eq.(\ref{eq:R.10}). The predicted variation with $\lambda$ (when $\lambda < 0$), in the slope of the scaling relation is also clearly apparent from Fig.\ref{fig:figR3}a.

\subsection{The expected value of $\lambda$ for increasing complexity \label{sec:lambdaTrend} }
It is natural to ask which of the above scenarios, $\lambda > 0$ or $\lambda < 0$, is appropriate for a real annotation process? A `least effort' argument would suggest that an annotator will attempt to limit the effort expended on annotating an object, and so $I$, the information retrieved per object, will most likely be fixed or only a weakly increasing function of $|{\mathcal O}|$. At fixed $z$ simple inspection of Eq.(\ref{eq:R.2b}) shows that $I$ is an increasing function of tree depth $L_{opt}$, and hence $|{\mathcal O}|$, for all values of $\lambda$. However, more detailed analysis of Eq.(\ref{eq:R.2b}) (in particular applying the Ratio test for series convergence, when $\lambda < 0$) reveals that as $L\rightarrow\infty$ we have (at fixed $z$),
\begin{eqnarray}
I & \sim & \frac{\log b}{2}L^{2}\;+\; O(L) \;\;\;,\;\; \lambda > 0 \;\; ,\nonumber \\
I & \rightarrow & {\rm Constant} \;\;\;,\;\;\lambda < 0 \;\; .
\end{eqnarray}

\noindent Therefore, we associate $\lambda < 0$ with a fixed, or only weakly growing value of $I$. This idea is corroborated if we consider the more intuitive scenario of a fixed value of $\bar{n}$ (as opposed to fixed fugacity $z$) and a fixed value of $I$, and determine the required value of $\lambda$ as the tree depth is $L$ is increased. Figure \ref{fig:figR3}b shows plots, against $L$, of the value of $\lambda$ required. The value of $\bar{n}$ is held fixed at $\bar{n} = 25/7$ and we have fixed $I(\lambda, L, \bar{n}) = c\times I(\lambda=0, L=5, \bar{n})$ for different values of $c$. The value of $\lambda$ is then obtained by simultaneous solution of Eq.(\ref{eq:R.2}) and Eq.(\ref{eq:R.2b}). Irrespective of the value of $c$ we can see that increasing $L$ at fixed $I$ and $\bar{n}$ leads to $\lambda <0$. Consequently, if the information or effort expended per object by an annotator is limited we expect small negative values of $\lambda$ to be the norm. Furthermore, we note that the amount of annotation information retrieved per object essentially determines the specificity with which the different classes of objects are discriminated. Thus, as data sets of increasing size will potentially sample an increasing number of the classes ${\mathcal P}$ present in the population, a fixed value of $I$ may not be sufficient to discriminate them, {\it i.e.}, the annotation data set will not be of adequate quality. Therefore, the value of $\lambda$ estimated from a real data set provides a topology derived metric to assess the quality of annotation and annotators, with a negative estimate for $\lambda$ giving possible cause for concern.

\section{ Discussion and Conclusions \label{sec:Conc}}

Although the real-world ontologies used in annotating data sets may be fixed in form, the annotation process, by nature of it often being a manual process, is subject to variation. We have used statistical mechanics to construct a formalism of that variability in the annotation process.
The formalism developed has allowed us to understand both the patterns seen in real data sets \cite{Kalankesh2012}, and to suggest measures of the ontology structure itself. This has been done by using a simple lattice-gas model of the term selection process combined with information theoretic concepts. Although measures of ontology structures \cite{Yao2011} already exist these do not tend to incorporate the effects of the expected variability in the annotation process. Likewise, many studies have previously combined information theoretic concepts with ontologies, but these works have largely not focused on assessing the underlying ontology structure. Instead studies have used information theoretic concepts  in assessing the similarity of individual terms within an ontology \cite{Resnik1995,Resnik1999, Calmet2004}, assessing the similarity between annotated objects \cite{Tao2007}, assessing the similarity between two annotation data sets \cite{Alterovitz2007}, assessing the similarity between users \cite{KohMui2001}, or in empirical studies of the growth in tag frequencies \cite{ChiMytkowicz2008}. Similarly, studies of annotation and tag statistics that relate to the underlying ontology structure exist within the physics literature \cite{Palla2008, Tibely2012}, but these have not been based upon a Hamiltonian model of the annotation process. By constructing a Hamiltonian model of the annotations we have built upon the work of Palla {\it et al.} \cite{Palla2008} and Tib\'{e}ly {\it et al.} \cite{Tibely2012}, and been able to construct, in a principled fashion, statistical mechanical measures of the ontology itself.

Having developed a simple lattice-gas model of the annotation process the statistical mechanical formalism has enabled us to progress further and to perform hypothetical experiments over an ensemble of different ontology structures, and in doing so gain new insight where it is not readily possible to perform real experiments. Our analysis of an ensemble of different ontology structures has focused on regular trees, though despite this restriction we still expect the high-level conclusions drawn in Sections \ref{sec:regTrees}, \ref{sec:simulation}, and \ref{sec:lambdaTrend} to be more universally valid. Firstly, our detailed analysis of $\beta\Omega^{*}$ identifies a natural or optimal ontology size, and associated growth scaling law, given the number of objects $|{\mathcal O}|$ to be annotated. The scaling laws derived provide us with 'rules-of-thumb' to decide when an ontology should be expanded to match the needs of the data sets to which it is being applied. Secondly, our analysis in Section \ref{sec:lambdaTrend} reveals a natural tendency towards  $\lambda <0$, as more complex ontologies are used to annotate larger collections of objects. This suggests that values of $\lambda$ appropriate to real-world annotation data sets will be typically small in magnitude, possibly even negative. This is borne out by the power-law exponents observed by Kalankesh {\it et al.} \cite{Kalankesh2012}. Small values of $\lambda$ would imply that annotators are typically recovering less information from the ontology than if they selected terms uniformly at random. As we equate effort expended with the information retrieved, a negative value of $\lambda$ suggests the effort being expended during the annotation is below that appropriate to the ontology complexity. This may be due to the structure of the ontology being more complex than is necessary to discriminate between the classes of objects being annotated, the annotation effort expended not being sufficient, or simply that there is currently insufficient evidence available to the annotators to discriminate between certain objects. With the statistical mechanical formalism we can potentially extend the analysis presented here to construct measures capable of distinguishing these different possibilities.

The statistical mechanical analysis we have presented here is based upon a relatively simple Hamiltonian. The richness of the behaviour we observe in the statistical mechanical model is more a consequence of the inhomogeneous field induced by the ontology topology. However, the statistical mechanical analysis we have presented here is far from complete. The advantage of having expressed the annotation process in terms of a statistical mechanical formalism is that we can easily extend our analysis to obtain quantitative results for other scenarios or other Hamiltonians. For example, to further reflect more realistic annotation patterns the analysis could be extended to take into account the non-independence of term usage. Indeed, term co-occurrence frequencies have been studied by Tibely {\it et al.} \cite{Tibely2012}, with terms that are closer (as measured by path length on the ontology DAG) occurring together more frequently than those further apart. Incorporating tag co-occurrence probabilities would turn our lattice-gas model from a non-interacting model to an interacting model. Alternatively, our statistical mechanical analysis can potentially be extended to account for more realistic aspects and nuances of the annotation process, {\it e.g.}, mis-selection of terms that are not appropriate to the object being annotated.

\appendix
\section{\label{sec:EulerMaclaurin}}
Applying the Euler-Maclaurin sum formula to $\sum_{l=0}^{L} \log \left ( 1 + zb^{l\lambda}\right )$ gives, to lowest order of expansion \cite{WhittakerWatson1990},
\begin{eqnarray}
&& \sum_{l=0}^{L} \log \left ( 1 + zb^{l\lambda}\right ) \; = \; \sum_{l=0}^{L} \log \left ( 1 + za^{l}\right ) \nonumber \\
 & = & L\int_{0}^{1}ds \log \left ( 1+za^{sL}\right )\;+\; \frac{1}{2}\log\left ( 1+za^{L}\right )
 \;+\; \frac{1}{2}\log\left ( 1+z\right ) \nonumber \\
 & + & \frac{1}{2}B_{2}\log a\left ( \frac{za^{L}}{1+za^{L}} \;-\;\frac{z}{1+z} \right ) \; - \; \frac{(\log a)^{2}}{2}\int_{0}^{L}ds B_{2}\left ( s-\lfloor s\rfloor\right )\frac{za^{s}}{(1+za^{s})^{2}}\;\; , \nonumber \\
\label{eq:A0.1}
\end{eqnarray}

\noindent where $B_{2}$ is the second Bernoulli number, $B_{2}(x) = x^{2} - x + \frac{1}{6} $ is the second Bernoulli polynomial, and $a=b^{\lambda}$. The first integral on the right-hand-side of Eq.(\ref{eq:A0.1}) can be expressed in terms of the dilogarithm function, $Li_{2}(x)$ \cite{Zagier2007, Kirillov1995}, as,
\begin{equation}
L\int_{0}^{1}ds \log \left ( 1+za^{sL}\right )\;=\; \frac{1}{\log a}\left ( Li_{2}(-z)\;-\;Li_{2}(-za^{L})\right )\;\; .
\label{eq:A0.2}
\end{equation}

\noindent With $\frac{za^{s}}{(1+za^{s})^{2}} > 0\;\forall s \in [0,L]$ we can bound the last term in Eq.(\ref{eq:A0.1}) simply by replacing $B_{2}(s-\lfloor s\rfloor)$ by its minimum and maximum values in the integrand ($-\frac{1}{12}$ and $\frac{1}{6}$, respectively). This yields bounds for
$\sum_{l=0}^{L} \log \left ( 1 + za^{l}\right )$ as,
\begin{eqnarray}
&&\frac{1}{2}\log \left( 1+za^{L}\right )\;+\; \frac{1}{2}\log \left ( 1 + z \right ) \;\leq\; \sum_{l=0}^{L} \log \left ( 1 + za^{l}\right ) -L\int_{0}^{1}ds \log \left ( 1+za^{sL}\right ) \nonumber \\
&\leq & \frac{1}{2}\log \left( 1+za^{L}\right )\;+\;\frac{1}{2}\log\left (1+z\right ) \;+\;\frac{\log a}{8}\frac{za^{L}}{1+za^{L}}\;-\;\frac{\log a}{8}\frac{z}{1+z}\;\; .
\label{eq:A0.3}
\end{eqnarray}

\noindent To arrive at a final approximation to $\sum_{l=0}^{L} \log \left ( 1 + za^{l}\right )$ that is convenient for further analysis we simply take some point between the two bounds in Eq.(\ref{eq:A0.3}). By replacing $B_{2}(s-\lfloor s\rfloor)$ by its average value of zero, in the last integral in Eq.(\ref{eq:A0.1}), we arrive at the approximation,
\begin{eqnarray}
\sum_{l=0}^{L} \log \left ( 1 + za^{l}\right ) & \simeq &  \frac{1}{\log a}\left ( Li_{2}(-z)\;-\;Li_{2}(-za^{L})\right ) \; + \; \frac{1}{2}\log \left( 1+za^{L}\right ) \nonumber \\
& + &\frac{1}{2}\log\left (1+z\right ) \; + \;\frac{\log a}{12}\left ( \frac{za^{L}}{1+za^{L}}\;-\;\frac{z}{1+z}\right )\;\; .
\label{eq:A0.4}
\end{eqnarray}

\noindent We should note that the approximation in Eq.(\ref{eq:A0.4}) is simply equivalent to dropping the remainder term in the representation of the Euler-Maclaurin sum formula we have used.

\begin{figure*}
\begin{center}
\scalebox{0.23}{%
\includegraphics*[bb = 0 0 792 612]{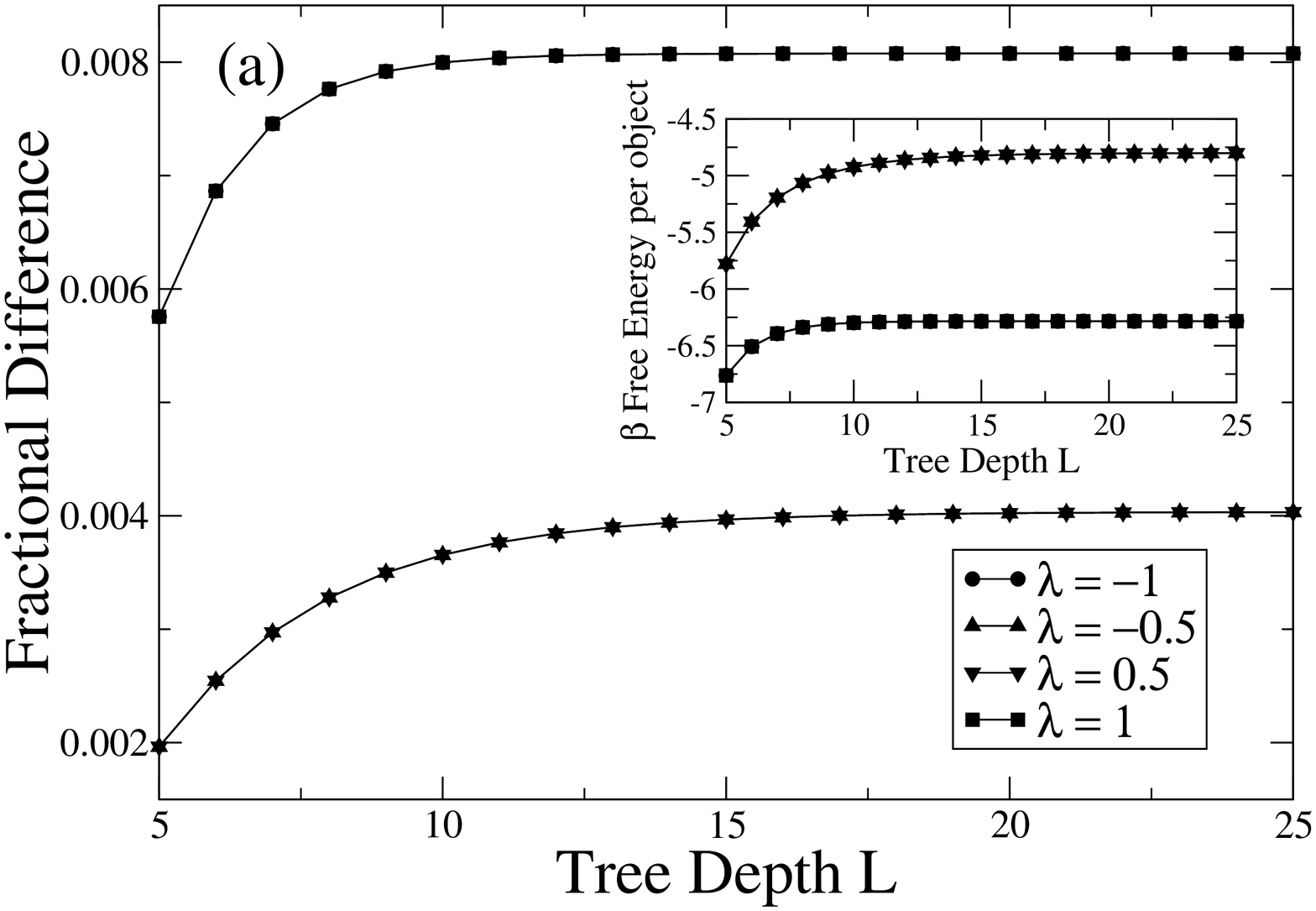}
\includegraphics*[bb = 0 0 792 612]{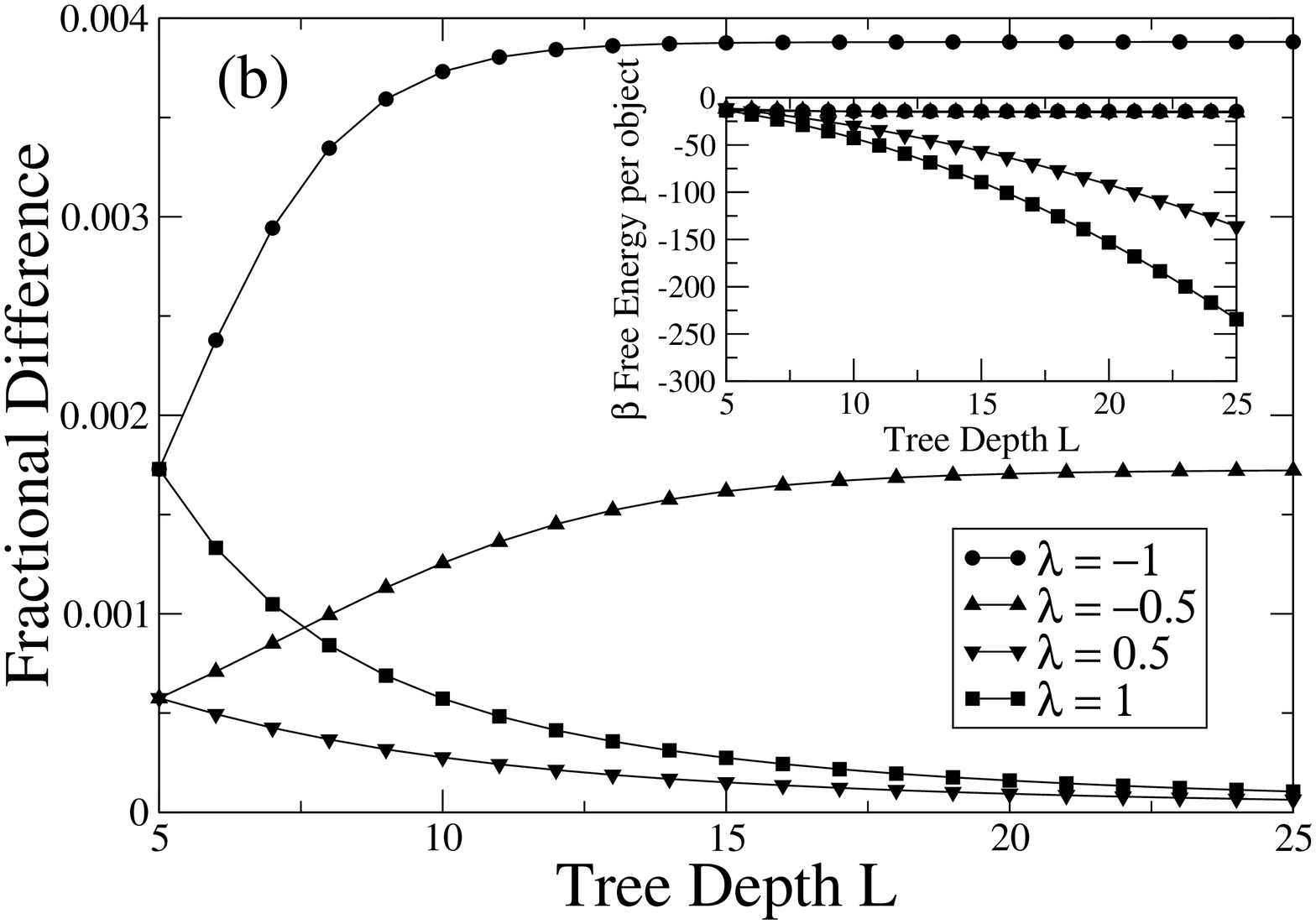}}
\end{center}
\caption{Plots versus tree depth $L$, at different values of $\lambda$, of the fractional difference between the exact free energy for a regular tree and the free energy calculated using the approximation Eq.(\ref{eq:A0.4}). The inset shows the exact free energy values. In all plots we have set the number of objects $|{\mathcal O}| = 1000$. In a) the number of annotation terms per object, $\bar{n}$, has been fixed at $\bar{n} = 25/7$, whilst in b) the fugacity $z$ has been fixed to achieve a value of $\bar{n} = 25/7$ when $L=5$. }
\label{fig:figR1}
\end{figure*}

The approximation developed in Eq.(\ref{eq:A0.4}) is in fact extremely accurate for the values of $z,a$ and $L$ we are interested in. Figure \ref{fig:figR1} shows the fractional difference between the exact evaluation of the sum on the left-hand-side of Eq.(\ref{eq:A0.4}) and the approximation on the right-hand-side of Eq.(\ref{eq:A0.4}). Fig.\ref{fig:figR1}a shows the fractional difference for different values of $L$ and $\lambda$ at a fixed values of $\bar{n}$ and suggests the approximation in Eq.(\ref{eq:A0.4}) is typically accurate to within 1 part in 100 for fixed $\bar{n}$. For the plots in Fig.\ref{fig:figR1}a the same value for the fugacity $z$ has been used in both free energy calculations. Namely, we have used the value of $z$ required to achieve a value of $\bar{n} = 25/7$, as determined from the exact summation relation in Eq.(\ref{eq:R.2}). Similarly, in Fig.\ref{fig:figR1}b shows the fractional difference between the exact summation in Eq.(\ref{eq:A0.4}) and the integral based approximation, at different values of $L$ and $\lambda$ but for a fixed value of $z$. Again, the same fixed value for the fugacity $z$ has been used in both free energy calculations, namely that required to achieve a value of $\bar{n} = 25/7$ when $L=5$, as determined from exact summation relation in Eq.(\ref{eq:R.2}). Fig.\ref{fig:figR1}b would suggest that for fixed $z$ the approximation in Eq.(\ref{eq:A0.4}) is again accurate to 2 parts in 1000.

\section{\label{sec:proof1}}
To observe growth in $L_{opt}$ as $|{\mathcal O}| \rightarrow\infty$, at fixed $\bar{n}$ with $\lambda < 0$, we require,
\begin{equation}
z^{2}_{0}\frac{f^{'}_{1}(0)f_{2}^{''}(z_{0})}{f_{2}^{'}(z_{0}) + z_{0}f_{2}^{''}(z_{0})} \; < \; 0\;\;.
\label{eq:A1.0}
\end{equation}

\noindent We start by analysing the behaviour of the numerator of Eq.(\ref{eq:A1.0}). From Eq.(\ref{eq:R.3e}) and Eq.(\ref{eq:R.3f2}) we have that $f_{1}(x)$ is defined as,
\begin{equation}
f_{1}(x)\;=\;-\frac{1}{\log a}Li_{2}(-x)\;+\; \frac{1}{2}\log ( 1+x )\; + \;\frac{\log a}{12}\frac{x}{1+x} \;\;,
\label{eq:A1.1}
\end{equation}

\noindent from which we have,
\begin{equation}
f_{1}^{'}(0)\;=\;\frac{1}{\log a}\;+\;\frac{1}{2}\;+\;\frac{\log a}{12}\;\;.
\label{eq:A1.2}
\end{equation}

\noindent We wish to determine the behaviour of $f_{1}^{'}(0)$ for $a\in\field{R}$ and $0 < a < 1$. To do so we seek the roots of $f_{1}^{'}(0) = 0$ as a function of $y=\log a < 0$. That is we solve,
\begin{eqnarray}
& & \frac{1}{y} \;+\;\frac{1}{2}\;+\;\frac{y}{12} = 0 \;\;\;\;,\;\;y\;=\;\log a \nonumber \\
&\Rightarrow& y^{2}\;+\;6y\;+\;12\;=\; 0\;\;.
\label{eq:A1.3}
\end{eqnarray}

\noindent As the discriminant of this quadratic equation is -12, there are no real roots. Therefore $f_{1}^{'}(0)$ maintains the same sign for all real values of $y <0$. It is a simple matter to confirm that $f_{1}^{'}(0) = -\frac{7}{12}$ at $y=-1$ and so $f_{1}^{'}(0) < 0$ for all $\lambda < 0$ with $b > 1$.

From Eq.(\ref{eq:R.3e}) and Eq.(\ref{eq:R.3f2}) we have that $f_{2}(x)$ is defined as,
\begin{equation}
f_{2}(x)\;=\;\frac{1}{\log a}Li_{2}(-x)\;+\; \frac{1}{2}\log ( 1+x )\; - \;\frac{\log a}{12}\frac{x}{1+x} \;\;,
\label{eq:A1.4}
\end{equation}

\noindent from which we find,
\begin{eqnarray}
f^{''}_{2}(x) & = & \frac{1}{\log a}\left ( \frac{1}{x^{2}}\log(1+x)\;-\; \frac{1}{x(1+x)}\right ) \;-\;\frac{1}{2}\frac{1}{(1+x)^{2}}\;+\;\frac{\log a}{6}\frac{1}{(1+x)^{3}}\;\;. \nonumber \\
&&
\label{eq:A1.5}
\end{eqnarray}

\noindent With $\log a < 0$ it is clear that if $\log(1+x) > x/(1+x)\;, \forall x >0$ then we will have $f^{''}_{2}(x) < 0\;,\forall x > 0$. At $x=0$ we have $\log(1+x) = x/(1+x)$. We also have,

\begin{eqnarray}
&& \frac{d\log(1+x)}{dx}\;=\; \frac{1}{1+x}\;\;\;\;\;,\;\;\;\;\;\frac{d [x/(1+x)]}{dx}\;=\;\frac{1}{(1+x)^{2}} \nonumber \\
&\Rightarrow&  \frac{d\log(1+x)}{dx}\; \ge\; \frac{d [x/(1+x)]}{dx}\;\;\;\;\forall x \ge 0\;\;, \nonumber \\
&\Rightarrow&  \log(1+x)\; \ge\; \frac{x}{1+x}\;\;\;\;\forall x \ge 0\;\;,
\label{eq:A1.6}
\end{eqnarray}

\noindent With $z^{2}_{0} > 0$ we find that the numerator of Eq.(\ref{eq:A1.0}) is negative.\newline

\noindent Now considering the behaviour of the denominator of Eq.(\ref{eq:A1.0}) we have,
\begin{equation}
f_{2}^{'}(z_{0}) + z_{0}f_{2}^{''}(z_{0})\;=\;\frac{d\bar{n}(z_{0})}{dz_{0}}\;\;.
\label{eq:A1.7}
\end{equation}

\noindent For an exactly calculated partition function we might expect $\bar{n}(z_{0})$ to be monotonically increasing in $z_{0}$. However, as we are utilising an integral approximation to the exact expression for $\beta\Omega^{*}$ given in Eq.(\ref{eq:R.1}), we continue our analysis with $d\bar{n}(z_{0}) / dz_{0}$ as given above. Explicitly, we find,
\begin{equation}
f_{2}^{'}(x) + xf_{2}^{''}(x)\;=\;-\frac{\left [ (12-6y+y^{2})\;+\;x(24-6y-y^{2})\;+\;12x^{2}\right ]}{12y(1+x)^{3}}\;\; .
\label{eq:A1.8}
\end{equation}

\noindent where again $y=\log a < 0$. Examination of the discriminant (with respect to $x$) of the numerator in the right-hand-side of Eq.(\ref{eq:A1.8}) reveals that in order for the right-hand-side of Eq.(\ref{eq:A1.8}) to be positive requires $\log a < -6 -4\sqrt{6}\simeq -15.8$. Therefore, overall we find that the left-hand-side of Eq.(\ref{eq:A1.0}) is negative for $ 0 > \log a > -6 -4\sqrt{6}$. This threshold value of $a$ we regard as somewhat extreme and outside the range that we are likely to be interested in - a value of $\log a = -15.8$ corresponds to a reduction by a factor of more than $10^{7}$ in the probability of selecting a term, in going from one layer of the tree to the next layer. The possibility that $d\bar{n}(z) / dz$ may be negative for some $z$ in the limit $L\rightarrow\infty$ is most likely due to a breakdown in the accuracy of the integral approximation in Eq.(\ref{eq:A0.1}). Indeed, from Eq.(\ref{eq:R.2}) we see that the exact expression for $d\bar{n}(z) / dz$ is given by,
\begin{equation}
\frac{d\bar{n}}{dz}\;=\;\frac{1}{z}\sum_{l=0}^{L}\frac{1}{(1+za^{l})^{2}}\;\;,
\label{eq:A1.9}
\end{equation}

\noindent which is clearly positive for all $\lambda$ and for all $z > 0 , a > 0$.

\section{\label{sec:proof2}}
For fixed $\bar{n}$ and $\lambda > 0$ the leading order contribution to $\beta\Omega^{*}$ from the free energy, that is dependent upon $L$, is given by,
\begin{equation}
|{\mathcal O}|a^{-L}\hat{z}_{1}\hat{z}_{2}f^{''}_{1}(\hat{z}_{1})\;\;.
\label{eq:A2.1}
\end{equation}

\noindent Using the relations in Eq.(\ref{eq:R.14}) we can re-write $\hat{z}_{1}\hat{z}_{2}f^{''}_{1}(\hat{z}_{1})$ as,
\begin{equation}
\hat{z}_{1}\hat{z}_{2}f^{''}_{1}(\hat{z}_{1})\;=\; -\frac{\hat{z}_{1}^{2}f^{''}_{1}(\hat{z}_{1})f^{'}_{2}(0)}{f^{'}_{1}(\hat{z}_{1})+ \hat{z}_{1}f^{''}_{1}(\hat{z}_{1})}\;\;.
\label{eq:A2.2}
\end{equation}

\noindent From the definition of $f_{1}(x)$ given in Eq.(\ref{eq:A1.1}) we find that,
\begin{equation}
f^{'}_{1}(\hat{z}_{1})+ \hat{z}_{1}f^{''}_{1}(\hat{z}_{1}) \;=\;
\frac{1}{\log a}\frac{1}{1+\hat{z}_{1}}\;+\;\frac{1}{2}\frac{1}{(1+\hat{z}_{1})^{2}}\;+\;\frac{\log a}{6}\frac{2+\hat{z}_{1}}{(1+\hat{z}_{1})^{3}}\;\;.
\label{eq:A2.3}
\end{equation}

\noindent With $\hat{z}_{1}\ge 0$ and $a>1$ we find that the denominator in Eq.(\ref{eq:A2.2}) is positive. Of the three factors contributing to the numerator in Eq.(\ref{eq:A2.2}), $\hat{z}_{1}^{2}$ is clearly positive. The form for $f_{2}(x)$ is already defined in Eq.(\ref{eq:A1.4}), from which we find,
\begin{equation}
f^{'}_{2}(0)\;=\;-\frac{1}{\log a}\;+\;\frac{1}{2}\;-\;\frac{\log a}{12}\;\;.
\label{eq:A2.5}
\end{equation}

\noindent As with Eq.(\ref{eq:A1.2}) we find that $f^{'}_{2}(0)=0$ has no real roots and so has the same sign for all $\log a > 0$. At $\log a = 1$ we find $f^{'}_{2}(0)=-\frac{7}{12}$ and so conclude that $f^{'}_{2}(0) < 0\;, \forall \log a > 0$.

\noindent For the remaining factor in the numerator of Eq.(\ref{eq:A2.2}) we have,
\begin{eqnarray}
f^{''}_{1}(x) & = & \frac{1}{\log a}\left ( -\frac{1}{x^{2}}\log(1+x)\;+\; \frac{1}{x(1+x)}\right )\;-\; \frac{1}{2}\frac{1}{(1+x)^{2}}\;-\;\frac{\log a}{6}\frac{1}{(1+x)^{3}}\;\;. \nonumber \\
&&
\label{eq:A2.6}
\end{eqnarray}

\noindent With $\log a > 0$ it is clear that we will have $f^{''}_{1}(x) < 0\;,\forall x > 0$ if $\log(1+x) > x/(1+x)\;, \forall x >0$. This latter relationship has already been proved in Eq.(\ref{eq:A1.6}), from which we can ultimately conclude $f^{''}_{1}(\hat{z}_{1}) \le 0\,\; \forall \hat{z}_{1} \ge 0$. Combining the three inequalities above relating to the numerator and denominator of Eq.(\ref{eq:A2.2}) we finally conclude that,

\begin{equation}
\hat{z}_{1}\hat{z}_{2}f^{''}_{1}(\hat{z}_{1})\;=\; -\frac{\hat{z}_{1}^{2}f^{''}_{1}(\hat{z}_{1})f^{'}_{2}(0)}{f^{'}_{1}(\hat{z}_{1})+ \hat{z}_{1}f^{''}_{1}(\hat{z}_{1})}\;\le \; 0\;\;,\;\;\forall \hat{z}_{1} \ge 0\;\;.
\label{eq:A2.8}
\end{equation}

\bibliography{HoyleBrass_arxiv2016}
\bibliographystyle{unsrt}

\end{document}